\documentclass[twoside,draft]{article}

\usepackage{amsfonts}
\usepackage{stmaryrd}
\usepackage{amssymb}
\usepackage{euscript}
\usepackage{amsthm}
\usepackage{amsmath}
\usepackage{amscd}
\usepackage{latexsym}
\usepackage{mathrsfs}
\usepackage{graphicx}
\usepackage{color}
\usepackage{dsfont}
\usepackage{bm}

\newtheorem{theorem}{Theorem}[section]
\newtheorem{lemma}{Lemma}[section]
\newtheorem{proposition}{Proposition}[section]
\newtheorem{corollary}{Corollary}[section]
\newtheorem{remark}{Remark}[section]

\newtheorem{definition}{Definition}[section]
\newtheorem{assumption}{Assumption}[section]

\newcommand{\beqs}{\begin{eqnarray*}}
\newcommand{\eeqs}{\end{eqnarray*}}

\newcommand{\psup}{\overline{p}}
\newcommand{\pinf}{\underline{p}}
\newcommand{\pmin}{\breve{p}}

\def\proof{\noindent {\it  Proof. $\, $}}
\def\finproof{\hfill $\Box$} 

\newcommand{\Lll }{L}
\newcommand{\Lld }{L^d}
\newcommand{\Llc }{L^c}
\newcommand{\Uuu }{U}
\newcommand{\Uud }{U^d}
\newcommand{\Uuc }{U^c}

\newcommand{\cLll }{\ell}
\newcommand{\cLld }{\ell^d}
\newcommand{\cLlc }{\ell^c}
\newcommand{\cUuu }{u}
\newcommand{\cUud }{u^d}
\newcommand{\cUuc }{u^c}

\newcommand{\Xhh }{X^h}

\newcommand{\wt }{\widetilde }
\newcommand{\wh }{\widehat }

\newcommand{\sigmash }{\sigma^{*,h}}
\newcommand{\sigmasc }{\sigma^{*,c}}
\newcommand{\taush }{\tau^{*,h}}
\newcommand{\tausc }{\tau^{*,c}}

\newcommand{\Mg }{\cM^p(\cC^g)}

\newcommand{\cEgH}{\cE^{g,H}}
\newcommand{\hHh}{H}

\newcommand{\cEgh}{\cE^{g,h}}
\newcommand{\cEgc}{\cE^{g,c}}

\newcommand\I{\mathds{1}}

\newcommand\Xxb{X^b}

\newcommand\Vben{V^b}

\def\phi{\varphi }

\def\P{{\mathbb P}}

\newcommand{\GCC}{\operatorname{GCC}\,}

\def\cCg{{\mathcal C}^g}

\def\cB{{\mathcal B}}
\def\cC{{\mathcal C}}
\def\cT{{\mathcal T}}
\def\cG{{\mathcal G}}

\def\cE{{\mathcal E}}
\def\cM{{\mathcal M}}
\def\cS{{\mathcal S}}
\def\cV{{\mathcal V}}
\def\cK{{\mathcal K}}
\def\cP{{\mathcal P}}

\def\rr{{\mathbb R}}

\def\gg{{\mathbb G}}

\newcommand{\Keywords}[1]{\par\noindent{\small{\bf Keywords\/}: #1}}
\newcommand{\Class}[1]{\par\noindent{\small{\bf Mathematics Subjects Classification (2010)\/}: #1}}

\title{{\Large \bf ARBITRAGE-FREE PRICING OF GAME OPTIONS \\ IN NONLINEAR MARKETS} \vskip 45 pt }

\author{Edward Kim$\,^{b}$, Tianyang Nie$\,^{a}$\footnote{The research of T. Nie and M. Rutkowski was supported by the DVC Research Bridging Support Grant {\it Pricing of American and game options in markets with frictions}. The work of T. Nie was supported by the National Natural Science Foundation of China (No. 11601285) and the Natural Science Foundation of Shandong Province (No. ZR2016AQ13).} \ and Marek Rutkowski$\,^{b,c}$ \\ \\
\\$^{a\,}$School of Mathematics, Shandong University,\\ Jinan, Shandong
250100, China\\ \\  $^{b\,}$School of Mathematics and Statistics, University of Sydney
\\ Sydney, NSW 2006, Australia\\ \\ $^{c\,}$Faculty of Mathematics and Information Science,
Warsaw University of Technology, \\ 00-661 Warszawa, Poland \\ }

\date{\vskip 35 pt \today \vskip 30 pt}

\begin{document}

\maketitle

\begin{abstract}
The goal is to re-examine and extend the findings from the recent paper by Dumitrescu et al.~\cite{DQS2017} who studied game options within the nonlinear arbitrage-free pricing approach developed in El Karoui and Quenez~\cite{EQ1997}.
We consider the setup introduced in Kim et al.~\cite{KNR2018} where contracts of an American style were examined.
We give a detailed study of unilateral pricing, hedging and exercising problems for the counterparties within a general nonlinear setup. We also present a BSDE approach, which is used to obtain more explicit results under suitable assumptions about solutions to doubly reflected BSDEs.
\vskip 20 pt
\vskip 10 pt
\Class{91G40,$\,$60J28}
\end{abstract}

\newpage

%

\section{Introduction} \label{sec1}

Unlike contracts of an American style, the so-called {\it game options} are symmetric between the two counterparties, hereafter referred to as the {\it hedger} and the {\it counterparty}, in the sense that both parties have the right to stop the contract before its nominal maturity date, denoted as $T$. We adopt here the convention according to which a random moment when the game contract is stopped by the hedger is called a {\it cancellation time} whereas for the corresponding moment for the counterparty is called an {\t exercise time}. Within the framework of the classical Black-Scholes options pricing model, the notion of an abstract game option was first introduced and studied by Kifer~\cite{KY2000} who coined the term {\it Israeli option}. His results were subsequently extended to a general (possibly incomplete) semimartingale model by Kallsen and K\"uhn~\cite{KK2004}. Arbitrage-free pricing and rational cancellation/exercising of game options (in particular, convertible bonds) within the framework of a linear market model have been studied by several authors, to mention a few: Ayache et al.~\cite{A2003}, Bielecki et al.~\cite{BCJ2008}, Dolinsky and Kifer~\cite{DK2007}, Hamad\'ene \cite{H2007}, Kallsen and K\"uhn~\cite{KK2004,KK2005}, Matoussi et al.~\cite{MPP2014}, K\"uhn et al.~\cite{KKS2006} and Kyprianou~\cite{KYP2004}.

It is well known that valuation of game options, at least within the framework of a linear market model, is closely
related to the concept of the zero-sum two-person Dynkin stopping game, as first introduced by Dynkin~\cite{DY1969} and later modified by Neveu~\cite{N1975}. From the mathematical perspective, one should also mention papers on Dynkin games and doubly reflected backward stochastic differential equations (DRBSDEs) that underpin valuation of game contract, in particular, Bayraktar and Yao~\cite{BY2015}, Cr\'epey and Matoussi~\cite{CM2008}, Cvitani\'c and Karatzas~\cite{CK1996}, Dumitrescu et al.~\cite{DQS2016}, Essaky and Hassani~\cite{EH2013}, Grigorova et al.~\cite{GIOQ2017}, Grigorova and Quenez~\cite{GQ2017}, Hamad\`ene and Ouknine~\cite{HO2016}, and Lepeltier and San Mart\'in~\cite{LSM2004}. The interested reader is referred to Kifer~\cite{KY2013} for a recent survey of results on Dynkin games and their applications to game options.

More recently, some authors have studied nonlinear variants of a Dynkin game and their applications to game options.
The aim of this work is to re-examine and extend the findings from Dumitrescu et al.~\cite{DQS2017} (see also \cite{DQS2018} for the case of
American options) who have studied game options within the framework of a particular imperfect market model with default using the nonlinear arbitrage-free pricing approach developed in El Karoui and Quenez~\cite{EQ1997}. In contrast to \cite{DQS2017}, we place ourselves within the setup of a general nonlinear arbitrage-free market, as introduced in Bielecki et al.~\cite{BCR2018,BR2015}, and we examine general properties of unilateral fair prices for the two counterparties.

Let us introduce the notation for a general nonlinear market model. We consider a filtered probability space $(\Omega, \cG, \gg,\P )$ satisfying the usual conditions of right-continuity and completeness, where the filtration $\gg=(\cG_t)_{t \in [0,T]}$ models the flow of information available to all traders. For convenience, the inception date of a contract under consideration is set to be 0 and it is assumed that the initial $\sigma$-field $\cG_0$ is trivial. Moreover, all processes introduced in what follows are implicitly assumed to be $\gg$-adapted and, as usual, any semimartingale is assumed to be c\`adl\`ag.
For the sake of notational convenience, we assume throughout that trading conditions are identical for the two counterparties, although this assumption can be relaxed without any difficulty by modifying the notation. By convention, all cash flows of a contract are described from the perspective of the hedger. Hence when a cash flow is positive for the hedger, then the cash amount is paid by the counterparty and received by the hedger. Obviously, if a cash flow is negative for the hedger, then the cash amount is transferred from the hedger to the counterparty.  Let us state a formal definition of a game contract.

\begin{definition} \label{def.GCC} \rm
A {\it game contract} with $\gg$-adapted, c\`adl\`ag processes $A,X^h,X^c,\Xxb$ is a contract between the two parties,
called the {\it hedger} and the {\it counterparty}, where: \\
\noindent (i) $\sigma\in\cT$ is chosen by the hedger and called the {\it cancellation time},\\
\noindent (ii) $\tau\in\cT$ is chosen by the counterparty and called the {\it exercise time}. \\
The process $A$ with $A_0=0$ gives the cumulative cash flows of the contract up to its effective maturity $\sigma \wedge \tau \wedge T$.  The contract expires at time $\sigma \wedge \tau \wedge T$ and its terminal payoff at time $\sigma \wedge \tau \wedge T$,
as seen from the perspective of the hedger, is given by the following expression, for every $\sigma,\tau\in\cT$,
\begin{equation}   \label{defii}
I(X^h,X^c,\Xxb;\sigma,\tau):=X^h_\sigma\I_{\{\sigma<\tau\}}+X^c_{\tau}\I_{\{\tau<\sigma\}}+\Xxb_{\sigma}\I_{\{\tau=\sigma\}}.
\end{equation}
\end{definition}

For brevity, a  game contract is denoted either as $\GCC(A,X^h,X^c,\Xxb,\cT )$ or, simply, $\cCg$.
In the existing literature, it is common to assume that the inequalities $X^h_t < X^c_t$ and $X^h_t \le \Xxb_t \le X^c_t $ are satisfied for every $t \in [0,T]$. In that case, the payoff process $\Xxb$ in fact plays only a minor role, since one can prove that the solutions to the hedging and valuation problems for the hedger and the counterparty will depend only on the terminal value $\Xxb_T$ and thus the values $\Xxb_t$ for $t\in [0,T)$ are immaterial. We will make these assumptions in Section \ref{sec3} where we analyze a BSDE approach to game options, since they are convenient when dealing with solutions to DRBSDEs. This should be contrasted with the situation studied in
Section \ref{sec2}, where we do not need to make any assumptions about the ordering of the payoffs $X^h,X^c$ and $X^b$.

An important element of arbitrage-free pricing in a nonlinear market is the concept of the {\it benchmark wealth} $V^b(x)$, that is,
the process with respect to which arbitrage opportunities of a particular trader are quantified and assessed. As in Bielecki and Rutkowski~\cite{BR2015},  Bielecki et al.~\cite{BCR2018} and Kim et al.~\cite{KNR2018}, the benchmark wealth can be given by the equality $V^b(x)=V^0(x)$ where, for any initial endowment $x\in\rr$ of a trader, we set $V^0(x):=x B^{0,l}\I_{\{x \geq 0\}}+ x B^{0,b}\I_{\{x < 0\}}$
where the risk-free \textit{lending} (respectively, {\it borrowing}) \textit{cash account} $B^{0,l}$ (respectively, $B^{0,b}$) is used for unsecured lending (respectively, borrowing) of cash. Note that $V^0(x)$ represents the wealth process of a trader who decided at time 0 to keep his initial cash endowment $x$ in either the lending (when $x \geq 0$) or the borrowing (when $x<0$) cash account and who is not involved in any other trading activities between the times 0 and $T$. More generally, if a trader is endowed with an initial portfolio of assets (including also the savings account $B^{0,l}$ and $B^{0,b}$, which means that he can also be a borrower or a lender of cash) with the current market value $x$ at time 0 (ignoring bid-ask spreads and transaction costs), then $V^b(x)$ stands for the wealth process of the trader's static portfolio under the postulate that it is hold unchanged till the date $T$. By convention, the quantities $x_1$ and $x_2$ represent initial endowments of the hedger and the counterparty, respectively, and thus the processes $V^0(x_1)$ and $V^0(x_2)$ are their respective benchmark wealths.
From the economic perspective, the notion of the benchmark wealth can be seen as a plausible formalization of the well known idea of
{\it opportunity costs}, which is instrumental in financial decision making.  Note also that, even if we set $x_1=x_2=0$ and $V^b_t(0)=0$ for all $t \in [0,T]$, then, due to nonlinear dynamics of the wealth process,  unilateral prices will still be asymmetric. A particular choice for the benchmark process $V^b(x)$ is important in practical applications, but it is immaterial for all results presented in this work.

In Section \ref{sec2}, we work in an abstract nonlinear setup, meaning that we only make very general assumptions about the nonlinear dynamics of the wealth process of self-financing strategies. The main assumptions of that kind are the {\it forward monotonicity} and the  {\it strict forward monotonicity} of wealth processes (see Assumptions \ref{ass1.1w} and \ref{ass1.1s}, respectively), the {\it comparison} and the {\it strict comparison} postulates for trading strategies (Assumptions \ref{ass1.1bw} and \ref{ass1.1bs}, respectively), as well as the replicability postulates for the hedger and the counterparty (Assumptions \ref{ass1.1h} and \ref{ass1.1c}, respectively). The main result in Section \ref{sec2} is Theorem \ref{the4.4}  where we show that, under mild and natural assumptions about the underlying nonlinear market model, a unilateral {\it hedger's acceptable price} for the game contract is unique and corresponds to the fair price obtained through replication. To this end, one needs to properly define the concepts of a fair price and replication of a game contract in a nonlinear setup. Using the symmetric features of a game contract, we show in Theorem \ref{the4.5} that the same properties of pricing are true for the counterparty as well.

In Section \ref{sec3}, we present a BSDE approach to game options in a fairly general setup. We obtain there more explicit results regarding pricing, hedging and rational cancellation/exercise times using a BSDE approach without being specific about the dynamics of underlying assets, but by focusing instead on desirable properties of solutions to DRBSDEs. In Section \ref{sec3.4}, we show that the hedger's acceptable price can be characterized by a unique solution to the hedger's DRBSDE. Next, in Section \ref{sec3.4b}, we analyze
the sets of hedger's rational cancellation times and break-even times under the postulate that the contract is traded at time 0
at the hedger's acceptable price. The corresponding results for the counterparty are stated without proofs in Sections  \ref{sec3.4} and
\ref{sec3.4b}. It should be noted that the DRBSDEs for the hedger and the counterparty differ and thus, in general,
unilateral acceptable prices computed by the two counterparties will not coincide.

\newpage

\section{Unilateral Fair Pricing} \label{sec2}

Let $\cM=(\cB,\cS,\Psi)$ be a market model which is arbitrage-free with respect to European contracts in the sense of  Bielecki et al.~\cite{BCR2018,BR2015}. Here $\Psi$ stands for the class of all {\it admissible} trading strategies and $\Psi(y,D)$ denotes the class of all admissible trading strategies from $\Psi$ with the initial wealth $y\in\rr$ and with external cash flows $D$. For any trading strategy $\phi\in\Psi (y,D)$, we denote by $V(y,\phi,D)$ the {\it wealth process} of $\phi$. Obviously, the equality $V_0(y,\phi,D)=y$ holds for all $y\in\rr$ and any strategy $\phi$. It is assumed throughout that the processes $D,\Xhh,X^c$ and the wealth process $V(y,\phi,D)$ are c\`adl\`ag and $\gg$-adapted. We will gradually make more assumptions about the dynamics of wealth processes.

\subsection{Hedger's Fair Prices}   \label{sec2.1}

We first conduct a preliminary analysis of unilateral fair valuation from the hedger's viewpoint. Unlike in the case of options of an American style, which were studied by Kim et al.~\cite{KNR2018}, the symmetry of the game contract simplifies the analysis, in the sense that it suffices to address the valuation and hedging problem for the hedger and, subsequently, to apply analogous arguments to establish results pertaining to the counterparty.

We consider an extended market model, denoted as $\Mg$, in which a game contract $\cCg= \GCC(A,X^h,X^c,\Xxb,\cT)$ is traded at time~$0$ at some initial price $p$ where $p$ can be an arbitrary real number. We first give a preliminary analysis of unilateral fair valuation of a game contract by the hedger. We henceforth assume that the hedger (respectively, the counterparty) is endowed with an initial pre-trading wealth of $x_1\in\rr$ (respectively, $x_2\in\rr$) units of cash.

Note that, for brevity, the variables $(A,X^h,X^c,\Xxb)$ are frequently omitted in what follows when there is no danger of confusion; in particular, the payoff $I(X^h,X^c,\Xxb ;\sigma,\tau)$ will be usually denoted as $I(\sigma,\tau)$.
Similarly, since the process $A$ is fixed throughout, we will frequently write $V(x_1+p,\phi)$ instead of $V(x_1+p,\phi,A)$ when dealing with the hedger. By the same token, we will later write $V(x_2-p,\psi)$ instead of $V(x_2-p,\psi,-A)$ when examining trading strategies of the counterparty.

We introduce the following conditions regarding the presence (or absence) of unilateral hedger's gains with respect to his predetermined benchmark $\Vben (x_1)$.

\begin{definition} \label{def1.1x}
{\rm A quadruplet $(p,\phi,\sigma,\tau)\in\rr\times\Psi(x_1+p,A)\times\cT\times\cT$ is said to satisfy:
\[
\begin{array}[l]{llll}
&\text{(AO)}& \Longleftrightarrow &V_{\sigma \wedge \tau }(x_1+p,\phi)+I(\sigma,\tau)\geq \Vben_{\sigma\wedge\tau }(x_1) \\
& & & \text{\rm and } \P\big(V_{\sigma\wedge\tau}(x_1+p,\phi)+I(\sigma,\tau)>\Vben_{\sigma\wedge\tau }(x_1)\big)>0,\medskip \\
&\text{(SH)}& \Longleftrightarrow &V_{\sigma\wedge\tau}(x_1+p,\phi)+I(\sigma,\tau)\geq \Vben_{\sigma\wedge\tau }(x_1),\medskip \\
&\text{(BE)}&\Longleftrightarrow &V_{\sigma\wedge\tau}(x_1+p,\phi)+I(\sigma,\tau)=\Vben_{\sigma\wedge\tau}(x_1),\medskip \\
&\text{(NA)}&\Longleftrightarrow &\text{\rm either } V_{\sigma\wedge\tau}(x_1+p,\phi)+I(\sigma,\tau)=\Vben_{\sigma \wedge \tau }(x_1)\\
& & & \text{\rm or } \P\big(V_{\sigma\wedge\tau}(x_1+p,\phi)+I(\sigma,\tau)< \Vben_{\sigma\wedge\tau}(x_1)\big)>0.
\end{array}
\]}	
\end{definition}

Let us explain the meaning of acronyms appearing in Definition \ref{def1.1x}: (AO) stands for {\it arbitrage opportunity}, (SH) for  {\it superhedging}, (BE) for {\it break-even} and (NA) for {\it no-arbitrage}. For a detailed explanation of each property, see Definitions \ref{def4.3h}--\ref{def4.4y}.

\begin{definition} \label{def4.3h} {\rm
If condition (BE) is satisfied by a quadruplet  $(p,\phi ,\sigma,\tau)$, then a stopping time $\tau \in\cT $ is called a {\it hedger's break-even time} for $(p,\phi,\sigma )\in\rr\times\Psi (x_1+p,A) \times \cT$.}
\end{definition}


Property (SH) of $(p,\phi,\sigma,\tau)$ is referred to as the {\it hedger's superhedging at time $\tau$}. In view of the optional section theorem, it is easy to see that property (SH) holds for a given triplet $(p,\phi,\sigma)\in\rr\times\Psi(x_1+p,A)\times \cT$ and {\it all} $\tau\in\cT$ if and only if $V_{\sigma \wedge t}(x_1+p,\phi)+I(\sigma, t) \geq \Vben_{\sigma \wedge t}(x_1)$ for all $t \in [0,T]$.  This observation justifies the following definition of property (SH) for a triplet $(p,\phi,\sigma)$.

\begin{definition} \label{defssxx} { \rm
We say that a triplet $(p,\phi,\sigma)\in\rr\times\Psi(x_1+p,A)\times\cT$ satisfies condition (SH) if the inequality $V_{\sigma \wedge t}(x_1+p,\phi)+I(\sigma, t) \geq \Vben_{\sigma \wedge t}(x_1)$ holds for all $t \in [0,T]$.  In that case, a triplet $(p,\phi,\sigma)$ is called a {\it hedger's superhedging strategy} in the extended market $\Mg$.}
\end{definition}

Property (AO) is referred to as the {\it hedger's strict superhedging condition} (or the {\it hedger's arbitrage opportunity}) for $(p,\phi,\sigma)$ at time $\tau$.  Note that if a triplet $(p,\phi,\sigma)$ is such that the inequality $V_{\sigma \wedge t}(x_1+p,\phi)+I(\sigma, t)>\Vben_{\sigma \wedge t}(x_1)$ holds for every $t \in [0,T]$, then condition (AO) is satisfied by $(p,\phi,\sigma)$ for every $\tau\in\cT$, but the converse implication does not hold, in general.

\begin{definition} \label{def4.4}{ \rm
We say that a triplet $(p,\phi,\sigma)\in\rr\times\Psi(x_1+p,A)\times\cT$ satisfies condition (AO) if a quadruplet $(p,\phi,\sigma,\tau)$ complies with (AO) for all $\tau\in\cT$.  In that case, we also say that $(p,\phi,\sigma)$ creates a {\it hedger's arbitrage opportunity} in the extended market $\Mg$. We say that {\it no hedger's arbitrage arises for} $(p,\phi,\sigma)$ if there exists $\tau\in\cT$ such that the quadruplet $(p,\phi,\sigma,\tau)$ satisfies (NA).}
\end{definition}

\begin{definition} \label{def4.4y}
 {\rm We say that $p^{f,h}(x_1)=p^{f,h}(x_1,X^h,X^c,\Xxb)$ is a {\it hedger's fair price} for $\cCg$ if no hedger's arbitrage opportunity
$(p,\phi,\sigma)$ may arise in the extended market $\Mg$ when $p=p^{f,h}(x_1)$. The set of hedger's fair prices equals
\[
\cK^{f,h}(x_1):=\big\{ p\in \rr \,|\ \forall \, (\phi,\sigma)\in \Psi(x_1+p,A)\times \cT \,\exists \,\tau\in\cT\!:
\text{$(p,\phi,\sigma,\tau) \in $ (NA)} \big\}
\]
and the upper bound for the hedger's fair prices equals $\psup^{f,h}(x_1):=\sup \cK^{f,h}(x_1)$.
If the equality $ \psup^{f,h}(x_1)=\max \cK^{f,h}(x_1)$ is satisfied  (that is, if $\psup^{f,h}(x_1)\in \cK^{f,h}(x_1)$), then $\psup^{f,h}(x_1)$ is denoted as $\wh{p}^{f,h}(x_1)$ and called the {\it hedger's maximum fair price} for $\cCg$.}
\end{definition}

\begin{assumption} 
\label{ass1.1w}
{\rm The {\it forward monotonicity} of wealth holds meaning that for all $x, p\in\rr,\, \phi\in\Psi (x+p,A)$ and $p'>p$ (respectively, $p'<p$), there exists a trading strategy $\phi'\in\Psi (x+p',A)$ such that $V_t (x+p',\phi' ) \geq V_t (x+p,\phi)$
(respectively, $V_t (x+p',\phi' ) \leq V_t (x+p,\phi)$) for every $t \in [0,T]$.}
\end{assumption}

Under the postulate of the forward monotonicity of wealth, we obtain the interval structure of the fair prices set.

\begin{lemma}
Let Assumption \ref{ass1.1w} be satisfied. If $p\in \cK^{f,h}(x_1)$ then for any $p'<p$, we have that $p'\in \cK^{f,h}(x_1)$.  Therefore, if $\cK^{f,h}(x_1) \ne \emptyset$, then either $\cK^{f,h}(x_1)=(-\infty,\psup^{f,h}]=(-\infty,\wh{p}^{f,h}]$ or $\cK^{f,h}(x_1)=(-\infty,\psup^{f,h})$.
\end{lemma}

\proof
We argue by contradiction. If $\cK^{f,h}(x_1)=\emptyset$, then $\psup^{f,h}=- \infty $. Let us now consider the case where $\cK^{f,h}(x_1) \ne \emptyset$. Assume that $ p\in\cK^{f,h}(x_1)$ and a number $p'$ such that $p'< p$ is not an hedger's fair price. Then there exists $\phi'\in\Psi (x_1+p',A)$ such that $(p',\phi' ,\tau)$ satisfy (AO) for every $\tau\in\cT$. Consequently, by Assumption \ref{ass1.1w}, there exists $\phi \in\Psi (x_1+p,A)$ such that $(p,\phi ,\tau)$ satisfy (AO) for every $\tau\in\cT$. This clearly contradicts the assumption that $p$ belongs to $\cK^{f,h}(x_1)$ and thus we conclude that the asserted properties are valid.
\finproof

\subsection{Hedger's Superhedging Costs}  \label{sec2.2}

In the next step, we analyze the hedger's superhedging costs for the game contract.

\begin{definition} \rm
The lower bound for the hedger's strict superhedging costs equals
\[
\pinf^{a,h}(x_1)=\pinf^{a,h}(x_1,X^h,X^c,\Xxb):=\inf \cK^{a,h}(x_1)
\]
 where
\[
\cK^{a,h}(x_1):=\big\{p\in \rr \,|\ \exists \,(\phi,\sigma)\in\Psi(x_1+p,A)\times\cT\!:\text{$(p,\phi,\sigma)\in$ (AO)}\big\}.
\]
If the equality $\pinf^{a,h}(x_1):=\min \cK^{a,h}(x_1)$ holds, then $\pinf^{a,h}(x_1)$ is denoted as $\pmin^{a,h}(x_1)$ and called the {\it hedger's minimum strict superhedging cost} for $\cCg$.
\end{definition}

For conciseness, the variables $(x_1,X^h,X^c,\Xxb)$ or $x_1$ will be frequently suppressed when dealing with the hedger's valuation problem and thus we will write $\pinf^{a,h}$ instead of $\pinf^{a,h}(x_1,X^h,X^c,\Xxb)$ or $\pinf^{a,h}(x_1)$, etc.
Notice that $\cK^{a,h}(x_1)$ is the complement of $\cK^{f,h}(x_1)$ and thus either
\begin{equation} \label{firstnn}
\cK^{f,h}(x_1)=(-\infty,\wh{p}^{f,h}\,]\ \,\text{and}\ \,\cK^{a,h}(x_1)=(\pinf^{a,h},\infty)
\end{equation}
or
\begin{equation} \label{secondnn}
\cK^{f,h}(x_1)=(-\infty,\psup^{f,h}\,)\ \,\text{and}\ \,\cK^{a,h}(x_1)=[\,\pmin^{a,h},\infty).
\end{equation}

\begin{definition} {\rm
The lower bound for the hedger's superhedging costs equals
\[
\pinf^{s,h}(x_1)=\pinf^{s,h}(x_1,X^h,X^c,\Xxb):=\inf \cK^{s,h}(x_1)
\]
where
\[
\cK^{s,h}(x_1):=\big\{p\in\rr \,|\ \exists \, (\phi,\sigma)\in\Psi(x_1+p,A)\times\cT\!: \text{$(p,\phi,\sigma) \in $ (SH)} \big\}.
\]
If the equality $\pinf^{s,h}(x_1):=\min \cK^{s,h}(x_1)$ holds, then  $\pinf^{s,h}(x_1)$ is denoted as $\pmin^{s,h}(x_1)$ and called the {\it hedger's minimum superhedging cost} for $\cCg $.}
\end{definition}

Since $\cK^{a,h}(x_1)\subseteq \cK^{s,h}(x_1)$, we always have that $\pinf^{s,h} \leq \pinf^{a,h}$ but, in principle, it may occur that $\pinf^{s,h}<\pinf^{a,h}=\psup^{f,h}$. To avoid this awkward situation, we introduce Assumption \ref{ass1.1s}, which ensures that $\pinf^{s,h}=\pinf^{a,h}$ and thus also $\pinf^{s,h}=\psup^{f,h}$.

\begin{assumption} 
\label{ass1.1s}
{\rm  The {\it strict forward monotonicity} of wealth holds meaning that for all $x, p\in\rr,\, \phi\in\Psi(x+p,A)$ and $p'>p$  (respectively, $p' <p$), there exists a trading strategy $\phi'\in\Psi (x+p',A)$ such that $V_t(x+p',\phi')>V_t(x+p,\phi)$ (respectively, $V_t(x+p',\phi')<V_t(x+p,\phi)$) for every $t\in [0,T]$.}
\end{assumption}

Obviously, Assumption \ref{ass1.1s} of the strict forward monotonicity of wealth is stronger than Assumption \ref{ass1.1w} of forward monotonicity of wealth and thus the former encompasses the latter.

\begin{lemma} \label{lem.1}
If Assumption \ref{ass1.1s} is satisfied, then $\psup^{f,h}=\pinf^{s,h}=\pinf^{a,h}$.
\end{lemma}

\proof
Let us first assume that $\cK^{s,h}(x_1)\neq \emptyset$ so that $\pinf^{s,h}< \infty$. Since Assumption \ref{ass1.1s} holds,
it is clear that for any $p \in \cK^{s,h}(x_1)$ and arbitrary $\varepsilon>0$, there exists a strategy $\phi'\in\Psi (x_1+p+\varepsilon,A)$ such that condition (AO) is satisfied by the pair $(p+\varepsilon,\phi')$. This means that $p+\varepsilon$ belongs to $\cK^{a,h}(x_1)$ and thus $p+\varepsilon \geq  \pinf^{a,h}$. From the arbitrariness of $p\in\cK^{s,h}(x_1)$ and $\varepsilon>0$, we obtain the inequality $ \pinf^{s,h}\ge\pinf^{a,h}$. Since $ \pinf^{s,h}\leq\pinf^{a,h}$, it is clear that $\pinf^{s,h}=\pinf^{a,h}$. Using also (\ref{firstnn}) and (\ref{secondnn}), especially $\cK^{a,h}(x_1)$ is the complement of $\cK^{f,h}(x_1)$, we obtain the equality $\psup^{f,h}=\pinf^{a,h}$ and thus we conclude that $ \psup^{f,h}=\pinf^{s,h}=\pinf^{a,h}$. Let us now assume that $\cK^{s,h}(x_1)=\emptyset$. Then $\pinf^{s,h}=\pinf^{a,h}=\infty$ and, since $\cK^{f,h}(x_1)$ is the complement of $\cK^{a,h}(x_1)$, we see that $\psup^{f,h}=\infty$ as well.
\finproof

\subsection{Hedger's Replication Costs} \label{sec2.3}

The next step is to introduce the concept of a replication cost. Let $(p,\phi,\sigma)$ be a hedger's superhedging strategy such that he breaks even at $\sigma \wedge \tau $ when the exercise time $\tau $ is coincidentally chosen by his counterparty. Then we say that $(p,\phi,\sigma)$ is a {\it hedger's replicating strategy} for a game contract.

\begin{definition} \label{defrepx}  \rm
The lower bound for hedger's replication costs for $\cCg$ is given by the equality
\[
\pinf^{r,h}(x_1)=\pinf^{r,h}(x_1,X^h,X^c,\Xxb):=\inf \cK^{r,h}(x_1)
\]
 where
\begin{align*}
\cK^{r,h}(x_1):=\big\{  p\in\rr \,|\ \exists \, (\phi ,\sigma ,\tau)\in\Psi(x_1+p,A)\times\cT\times\cT\!:
\text{$(p,\phi ,\sigma ) \in $ (SH)}\ \& \ \text{$(p,\phi ,\sigma ,\tau) \in $ (BE)}  \big\}.
\end{align*}
If the equality $\pinf^{r,h}(x_1):=\min \cK^{r,h}(x_1)$ holds, then  $\pinf^{r,h}(x_1)$ is denoted as $\pmin^{r,h}(x_1)$ and called the {\it hedger's minimum replication cost} for $\cCg$.
\end{definition}

\begin{definition} {\rm
The lower bound for hedger's fair replication costs for $\cCg$ is given by the equality
\[
\pinf^{f,r,h}(x_1)=\pinf^{f,r,h}(x_1,X^h,X^c,\Xxb):=\inf \cK^{f,r,h}(x_1)
\]
 where
\begin{align*}
\cK^{f,r,h}(x_1):=\big\{p\in\rr \,|\ & \exists \,(\phi,\sigma,\tau)\in\Psi(x_1+p,A)\times\cT\times\cT\!:\;
\text{$(p,\phi,\sigma) \in $ (SH)}\ \& \  \text{$(p,\phi,\sigma,\tau) \in $ (BE);}\\
 &\forall \,(\phi',\sigma')\in \Psi(x_1+p,A)\times\cT,\,\exists \,\tau'\in\cT\!: \text{$(p,\phi',\sigma',\tau')\in$ (NA)} \big\}.
\end{align*}
If $\pinf^{f,r,h}(x_1):=\min \cK^{f,r,h}(x_1)$, then  $\pinf^{f,r,h}(x_1)$ is denoted as $\pmin^{f,r,h}(x_1)$ and called the {\it hedger's minimum fair replication cost} for $\cCg$.}
\end{definition}

It is readily seen that $\cK^{s,h}(x_1)\supseteq \cK^{r,h}(x_1) \supseteq \cK^{f,r,h}(x_1)=\cK^{f,h}(x_1) \cap \cK^{r,h}(x_1)$ and thus $\pinf^{s,h} \le \pinf^{r,h} \le \pinf^{f,r,h}$.

\begin{lemma} \label{lem.price}
Let Assumption \ref{ass1.1s} be satisfied.\\
(i) If $\cK^{f,r,h}(x_1) \ne \emptyset $, then $-\infty<\wh{p}^{f,h}=\pmin^{r,h}=\pmin^{f,r,h}=\pmin^{s,h}<+\infty$. \\
(ii) If $\cK^{r,h}(x_1)  \ne \emptyset $, then $\psup^{f,h}=\pinf^{s,h} \leq \pinf^{r,h} < \infty $.\\
(iii) If $\cK^{r,h}(x_1)=\emptyset $, then $\psup^{f,h}=\pinf^{s,h} \leq \pinf^{r,h} =\infty $.
\end{lemma}

\proof
We first prove part (i). It is clear that it in enough to show $\pinf^{f,r,h} \le \pinf^{s,h}$. For this purpose, we notice that $\cK^{f,r,h}(x_1) \subseteq \cK^{f,h}(x_1)$. Since $\cK^{f,r,h}(x_1) \ne \emptyset$, it follows that $\pinf^{f,r,h}<+\infty$ and $\pinf^{f,r,h} \leq \psup^{f,h}$. Consequently, the equalities $\psup^{f,h}=\pinf^{r,h}=\pinf^{f,r,h}=\pinf^{s,h}$ are valid. Moreover, from the inclusion $\cK^{f,r,h}(x_1)\subseteq  \cK^{f,h}(x_1)$ and the equalities $\sup \,\cK^{f,h}(x_1)=\psup^{f,h}=\pinf^{f,r,h}=\inf \,\cK^{f,r,h}(x_1)$, we deduce that for arbitrary $p_1,p_2\in\cK^{f,r,h}(x_1)$ and $p_3\in\cK^{f,h}(x_1)$ we have that $p_1=p_2\ge p_3$. This means that $\cK^{f,r,h}(x_1)$ is a singleton and its unique element is not less than any element of $\cK^{f,h}(x_1)$. Consequently, $\wh{p}^{f,h}$ and $\pmin^{f,r,h}$ are well defined and satisfy $-\infty<\wh{p}^{f,h}=\pmin^{f,r,h}<+\infty$. Furthermore,  $\cK^{f,r,h}(x_1) \subseteq  \cK^{r,h}(x_1)$ and thus the equality $\pmin^{f,r,h}=\pinf^{r,h}$ implies that $\pmin^{r,h}$ is well defined and is equal to $\pmin^{f,r,h}$. We conclude that $\wh{p}^{f,h}=\pmin^{r,h}=\pmin^{f,r,h}=\pmin^{s,h}$. This means, in particular, that $\cK^{f,h}(x_1)=(-\infty,\wh{p}^{f,h}]= (-\infty,\pmin^{r,h}]=(-\infty,\pmin^{f,r,h}]$. Statements (ii) and (iii) are obvious consequences of Lemma \ref{lem.1}.
\finproof

\subsection{Hedger's Acceptable Price} \label{sec2.4}

We will henceforth work under the following postulate of the (backward) comparison property for trading strategies.

\begin{assumption} 
\label{ass1.1bw} {\rm The {\it comparison property} for trading strategies holds meaning that for all $\tau\in\cT,\, x, p,p'\in\rr,\phi\in\Psi (x+p,A)$ and $\phi'\in\Psi (x+p',A)$, if the inequality $V_{\tau}(x+p',\phi') \geq V_{\tau}(x+p,\phi)$ holds, then $p'\geq p$.}
\end{assumption}

Under Assumption \ref{ass1.1bw}, if for some $\gg$-adapted process $X$ and a given stopping time $\theta \in\cT$ there exists a pair $(x,\phi)\in\rr\times\Psi (x,A)$ satisfying $V_{\theta }(x,\phi)=X_{\theta }$, then $x$ is unique and it is denoted by $\cE^h_0(X_{\theta})$. We set $X^l_t:=\Vben_t(x_1)-X^c_t,\,X^u_t:=\Vben_t(x_1)-X^h_t$ and $X^m_t:=\Vben_t(x_1)-\Xxb_t$ for every $t \in [0,T]$ and we denote
\begin{equation} \label{eq4.4h}
J(x_1,X^l,X^u,X^m,\sigma,\tau):=\Vben_{\sigma \wedge \tau }(x_1)-I(\sigma ,\tau)=X^l_{\tau}\I_{\{\tau<\sigma\}}+ X^u_\sigma\I_{\{\sigma<\tau\}}+ X^m_{\sigma}\I_{\{\tau=\sigma\}}.
\end{equation}
The {\it hedger's relative reward} $J(x_1,X^l,X^u,X^m,\sigma,\tau)$ will be henceforth denoted as $J^h(\sigma,\tau)$. Note that it is not postulated in Section  \ref{sec2} that $X^l \leq X^m \leq X^u$, although we will need to make this assumption later on when addressing the hedger's valuation problem using a BSDE approach studied in Section \ref{sec3}.

The following assumption is used to ensure that the quantity $\cE^h_0(J^h(\sigma,\tau))$ is well defined for all stopping
times $\sigma , \tau \in \cT$.

\begin{assumption} 
\label{ass1.1h} {\rm
We assume that the process $J^h$ is {\it replicable} for the hedger, in the sense that for a given $x_1\in\rr$ and every $\sigma ,\tau \in\cT$ there exists a pair $(p,\phi)\in \rr\times\Psi(x_1+p,A)$ such that $V_{\sigma \wedge \tau }(x_1+p,\phi)=J^h(\sigma,\tau)$.}
\end{assumption}

The following lemma is obvious and thus its proof is omitted.

\begin{lemma}
If Assumptions \ref{ass1.1bw} and \ref{ass1.1h} are satisfied, then the hedger's nonlinear evaluation
$(\sigma ,\tau ) \mapsto \cE^h_0(J^h (\sigma,\tau ))$ is well defined for
all $\sigma , \tau \in \cT$ and is unique.
\end{lemma}

\begin{definition} \label{def4.4xa} {\rm
We say that a triplet $(v^h_0(x_1),\sigmash ,\taush )\in\rr\times\cT\times\cT $ solves the {\it hedger's optimal replication problem} for $\cCg$ if $v^h_0(x_1)=\cE^h_0(J^h(\sigmash ,\taush ))- x_1$ where the stopping times $\sigmash$ and $\taush$ are such that}
\begin{equation} \label{eq4.4b}
\cE^h_0(J^h(\sigmash ,\taush ))=\min_{\sigma\in\cT }\,\max_{\tau\in\cT } \,\cE^h_0(J^h(\sigma ,\tau)).
\end{equation}
\end{definition}

\begin{assumption} 
\label{ass4.4bv}
{\rm The hedger's optimal replication problem for $\cCg$ has a solution $(v^h_0(x_1),\sigmash ,\taush )$. Furthermore, for $p^{*,h}=v^h_0(x_1)$ there exists a trading strategy $ \phi^*\in\Psi (x_1+p^{*,h},A)$ such that the triplet $(p^{*,h},\phi^* ,\sigmash )$ satisfies condition (SH) and the quadruplet $(p^{*,h},\phi^*,\sigmash ,\taush)$ complies with condition (BE), so that the set $\cK^{r,h}(x_1)$ is nonempty.}
\end{assumption}


\begin{remark}
{\rm Notice that the hedger's optimal replication problem does not hinge on the existence of a saddle point for the nonlinear Dynkin game associated with $\cE^h$ and $J^h$. To be more specific, we do not assume that the corresponding nonlinear Dynkin game has the {\it value}, in the sense that}
\[
\inf_{\sigma\in\cT}\,\sup_{\tau\in\cT }\,\cE^h_0(J^h(\sigma,\tau))=\sup_{\tau\in\cT}\,\inf_{\sigma\in\cT }\,\cE^h_0(J^h(\sigma,\tau)).
\]
\end{remark}

\begin{lemma} \label{lem4.4d}
If Assumptions \ref{ass1.1bw}--\ref{ass4.4bv} are met, then $\pinf^{s,h}(x_1) \geq v^h_{0}(x_1)$.
\end{lemma}

\proof
The assertion is trivial when $\cK^{s,h}(x_1)=\emptyset$ since then $\pinf^{s,h}(x_1)=\infty $. Let thus $p$ belong to $\cK^{s,h}(x_1)\neq \emptyset$. Then there exists a pair $(\phi ,\sigma )\in\Psi (x_1+p,A)\times\cT $ such that $(p,\phi ,\sigma )$ satisfies (SH) so that $V_{\sigma \wedge t}(x_1+p,\phi)\geq J^h(\sigma,t)$ for all $t\in[0,T]$. Consequently, for all $\tau\in\cT$, we have that $V_{\sigma \wedge \tau }(x_1+p,\phi)\geq J^h(\sigma,\tau)$ and, due to Assumption \ref{ass1.1bw}, this implies that for all $\tau\in\cT$, we have $x_1+p \geq \cE^h_0 ( J^h(\sigma,\tau))$. It is now easy to see
\[
x_1+p \geq \sup_{\tau\in\cT } \,\cE^h_0 ( J^h(\sigma,\tau)) \geq \inf_{\sigma\in\cT }\sup_{\tau\in\cT } \,\cE^h_0 ( J^h(\sigma,\tau)).
\]
Using equation \eqref{eq4.4b} and Assumption \ref{ass4.4bv}, we conclude that $p \geq v^h_{0}(x_1)$. Since $p\in\cK^{s,h}(x_1)$ was arbitrary, this yields the desired inequality.
\finproof

\begin{lemma} \label{lem4.4c}
If Assumptions \ref{ass1.1s}--\ref{ass4.4bv} are met, then $\psup^{f,h}(x_1) \leq v^h_{0}(x_1)$.
\end{lemma}

\proof
We argue by contradiction. Suppose that $\psup^{f,h}(x_1)>v^h_{0}(x_1)$. From Assumption \ref{ass4.4bv}, for $p^{*,h}=v^h_{0}(x_1)$ there exists a trading strategy $\phi^*\in\Psi (x_1+ p^{*,h},A)$ such that $V_{\sigmash \wedge \taush }(x_1+p^{*,h},\phi^*)=J(\sigmash,\taush)$ and the triplet $(p^{*,h},\phi^*,\sigmash )$ satisfies (SH). Consequently, in view of Assumption \ref{ass1.1s}, for any $p'>v^h_{0}(x_1)$ there exists $\phi'\in\Psi (x_1+p',A)$ such that $(p',\phi',\sigmash)$ is a hedger's strict superhedging strategy. This contradicts the assumption that $\psup^{f,h}(x_1,X^c,X^h,\Xxb)>v^h_{0}(x_1)$, since we have shown that if $p$ belongs to the interval $(v^h_{0}(x_1),\psup^{f,h}(x_1))$, then $p'\in\cK^{f,h}(x_1)\cap\cK^{a,h}(x_1)=\emptyset$ (recall that $\cK^{a,h}(x_1)$ is the complement of $\cK^{f,h}(x_1)$).
\finproof

The next assumption, which is manifestly stronger than Assumption \ref{ass1.1bw}, is crucial to ensure that the solution
$v^h_{0}(x_1)$ to the hedger's optimal replication problem yields the hedger's maximum fair price $\wh{p}^{f,h}(x_1)$.

\begin{assumption} 
\label{ass1.1bs}
{\rm  For all $\tau\in\cT,\, x,p,p'\in\rr,\phi\in\Psi (x+p,A)$ and $\phi'\in\Psi (x+p',A)$, if the inequality $V_{\tau}(x+p',\phi')\geq V_{\tau}(x+p,\phi)$ holds, then $p'\geq p$. If, in addition, $V_{\tau}(x+p',\phi')\ne V_{\tau}(x+p,\phi)$, then $p'>p$.
} \end{assumption}

The next result shows that, under Assumptions \ref{ass1.1s} and \ref{ass1.1h}--\ref{ass1.1bs}, the hedger's maximum fair price coincides with the solution $v^h_{0}(x_1)$ to the hedger's optimal replication problem introduced in Definition \ref{def4.4xa}. It is useful to observe that
\begin{equation} \label{eq4.4bb}
\cE^h_0(J^h(\sigma,\tau))-x_1=\big\{p\in\rr\,|\ \exists\,\phi\in\Psi (x_1+p,A)\!:V_{\sigma\wedge\tau}(x_1+p,\phi)=J^h(\sigma,\tau)\big\},
\end{equation}
where, in view of Assumption \ref{ass1.1bs}, the set in the right-hand side has only one element.

\begin{theorem} \label{the4.4}
Let Assumptions \ref{ass1.1s} and \ref{ass1.1h}--\ref{ass1.1bs} be valid. If $p\in\cK^{a,h}(x_1)$, then the inequality $p>v^h_{0}(x_1)$ holds and thus $\wh{p}^{f,h}(x_1)=v^h_{0}(x_1)$. Moreover, $\cK^{f,r,h}(x_1)\ne\emptyset$ and
\begin{equation}  \label{eq4.4f}
v^h_{0}(x_1)=\wh{p}^{f,h}(x_1)=\pmin^{f,r,h}(x_1)=\pmin^{s,h}(x_1).
\end{equation}
The stopping time $\taush$ is a hedger's break-even time for the triplet $(\pmin^{f,r,h}(x_1),\phi^{f,r,h},\sigmash)$ where a
trading strategy $\phi = \phi^{f,r,h}$ is implicitly given by equation \eqref{eq4.4bb} with
$(p,\sigma,\tau) = (\pmin^{f,r,h}(x_1),\sigmash,\taush )$.
More explicitly, a trading strategy $\phi^{f,r,h}$ belongs to $\Psi ( x_1+ \pmin^{f,r,h}(x_1), A)$ and is such that
\[
V_{\sigmash \wedge\taush }\big(x_1+\pmin^{f,r,h}(x_1) ,\phi^{f,r,h}\big)=J^h\big(\sigmash,\taush\big).
\]
\end{theorem}

\proof
We denote (noticing Assumption \ref{ass1.1bw}, the right side of the following equation is singleton)
\begin{equation}  \label{eq4.4x}
p^{r,h}(x_1,\sigma ,\taush ):=\big\{p\in\rr\,|\ \exists\,\phi\in \Psi (x_1+ p,A)\!:\text{$(p,\phi,\sigma,\taush )\in $ (BE)}\big\}.
\end{equation}
Let $p$ be an arbitrary number from $\cK^{a,h}(x_1)$. Then there exists a pair $( \phi , \sigma ) \in\Psi (x_1+p,A)\times \cT$ such that for every $\tau\in\cT$ we have $V_{\sigma \wedge \tau }(x_1+p,\phi) \geq J^h(\sigma,\tau)$ and $\P\big(V_{\sigma\wedge\tau}(x_1+p,\phi)> J^h(\sigma,\tau)\big)>0$. Hence we obtain for $\tau=\taush$
\[
V_{\sigma \wedge \taush }(x_1+p,\phi) \geq J^h(\sigma,\taush),\quad \P\big(V_{\sigma\wedge\taush}(x_1+p,\phi)> J^h(\sigma,\taush )\big)>0.
\]
From equation \eqref{eq4.4x} and Assumption \ref{ass1.1bs}, we get
\[
V_0(x_1+p,\phi)=x_1+p>x_1+p^{r,h}(x_1,\sigma,\taush) ,
\]
but, in view of Assumption \ref{ass4.4bv}, we also have that
\[
x_1+ v^h_{0}(x_1)=\cE^h_0(J^h(\sigmash ,\taush ))\leq \cE^h_0(J^h(\sigma,\taush))= x_1+p^{r,h}(x_1,\sigma,\taush)
\]
and thus $p> v^h_{0}(x_1)$. From equation \eqref{firstnn} and Lemma \ref{lem4.4c}, we deduce that $v^h_{0}(x_1)=\wh{p}^{f,h}(x_1)$. Moreover, in view of Assumption \ref{ass4.4bv}, we also have that $v^h_{0}(x_1)\in\cK^{f,r,h}(x_1)$ and thus $\cK^{f,r,h}(x_1)\ne\emptyset $. Hence to establish \eqref{eq4.4f} it suffices to make use of Lemma \ref{lem.price}. The last part of the statement is an immediate consequence of Definitions  \ref{def4.3h} and \ref{def4.4xa}.
\finproof

The following definition hinges on Theorem \ref{the4.4}.

\begin{definition} \label{def1.17h}
{\rm If the set $\cK^{f,r,h}(x_1)$ is a singleton, then its unique element is denoted as $p^h(x_1)$ and called the
{\it hedger's acceptable price} for $\cCg$.}
\end{definition}
\begin{remark}\label{remark for the hedger's acceptable price}{\rm From Theorem \ref{the4.4} and the proof of Lemma \ref{lem.price}, we know that under Assumptions \ref{ass1.1s} and \ref{ass1.1h}--\ref{ass1.1bs}, the
 hedger's acceptable price for $\cCg$ is well defined.}
\end{remark}
\subsection{Counterparty's Acceptable Price}  \label{sec2.5}

Due to symmetric features of a game contract, to address the pricing problem for the counterparty, it suffices to make appropriate modifications in the statements of results from Section \ref{sec2.4}. Note, in particular, that the hedger is receiving the initial price $p$, whereas the counterparty is paying the price provided, of course, if $p$ is nonnegative. More formally, a number $p$ is added to the initial endowment of the hedger, but it is subtracted from the initial endowment of the counterparty. For this reason, the counterparty searches for maximum superhedging and replication costs and minimum fair prices, which are denoted as $\wh{p}^{s,c}(x_2),\,\wh{p}^{r,c}(x_2)$ and $\pmin^{f,c}(x_2)$, respectively.

\begin{definition} \label{defcou}
{\rm A quadruplet $(p,\psi,\sigma,\tau)\in\rr\times\Psi(x_2-p,-A)\times\cT\times\cT$ is said to satisfy
\[
\begin{array}[l]{llll}
&\text{(AO$'$)}& \Longleftrightarrow &V_{\sigma \wedge \tau }(x_2-p,\psi)-I(\sigma,\tau)\geq \Vben_{\sigma\wedge\tau }(x_2) \\
& & & \text{\rm and } \P\big(V_{\sigma\wedge\tau}(x_2-p,\psi)-I(\sigma,\tau)>\Vben_{\sigma\wedge\tau }(x_2)\big)>0,\medskip \\
&\text{(SH$'$)}& \Longleftrightarrow &V_{\sigma\wedge\tau}(x_2-p,\psi)-I(\sigma,\tau)\geq \Vben_{\sigma\wedge\tau }(x_2),\medskip \\
&\text{(BE$'$)}&\Longleftrightarrow &V_{\sigma\wedge\tau}(x_2-p,\psi)-I(\sigma,\tau)=\Vben_{\sigma\wedge\tau}(x_2),\medskip \\
&\text{(NA$'$)}&\Longleftrightarrow &\text{\rm either } V_{\sigma\wedge\tau}(x_2-p,\psi)-I(\sigma,\tau)=\Vben_{\sigma \wedge \tau }(x_2)\\
& & & \text{\rm or } \P\big(V_{\sigma\wedge\tau}(x_2-p,\psi)-I(\sigma,\tau)< \Vben_{\sigma\wedge\tau}(x_2)\big)>0.
\end{array}
\]}	
\end{definition}

Of course, all other definitions and results formulated and established in Sections \ref{sec2.1}-\ref{sec2.4}
for the hedger have analogous (although not identical) versions for the counterparty. Since there is no need to present
all of them here, we only state the following important definitions.

\begin{definition} \label{def4.3} {\rm
If condition (BE$'$) is satisfied by  $(p,\psi ,\sigma,\tau)$, then a stopping time $\sigma \in\cT $ is called a {\it counterparty's break-even time} for $(p,\psi,\tau )\in\rr\times\Psi (x_2-p,-A)$.}
\end{definition}

\begin{definition} \label{defsscx} { \rm
We say that a triplet $(p,\psi,\tau )\in\rr\times\Psi(x_2-p,-A)\times\cT$ satisfies condition (SH$'$) if the inequality $V_{\tau \wedge t}(x_2-p,\psi)- I(\tau , t) \geq \Vben_{\tau \wedge t}(x_2)$ holds for all $t \in [0,T]$.  In that case, a triplet $(p,\psi,\tau)$ is called a {\it counterparty's superhedging strategy} in the extended market $\Mg$.}
\end{definition}

We will identify conditions under which the counterparty's maximum fair replication cost, denoted as $\wh{p}^{f,r,c}(x_2)$, is well defined and coincides with a solution to the counterparty's optimal replication problem given by Definition \ref{def4.5xa}.
It is convenient to define 
$x^u_t:=X^c_t+\Vben_t(x_2),\, x^l_t:=X^h_t+\Vben_t(x_2)$ and $x^m_t:=\Xxb_t+\Vben_t(x_2)$ for every $t \in [0,T]$ and we denote
\begin{align} \label{eq4.5h}
\wt{J}(x_2,x^l,x^u,x^m,\sigma,\tau)&:= I(\sigma ,\tau)+\Vben_{\sigma\wedge\tau }(x_2) = J(x_2,x^u,x^l,x^m,\sigma,\tau) \nonumber \\
&=x^l_\sigma\I_{\{\sigma<\tau\}}+ x^u_{\tau}\I_{\{\tau<\sigma\}}+x^m_{\sigma}\I_{\{\tau=\sigma\}}.
\end{align}
For brevity, the {\it counterparty's relative reward} $\wt{J}(x_2,x^l,x^u,x^m,\sigma,\tau)$ is also denoted as $J^c (\sigma,\tau)$.

To formulate the counterparty's optimal replication problem, we postulate that Assumptions  \ref{ass1.1s} and \ref{ass1.1bw} hold but with the process $A$ replaced by $-A$. For a given process $Y$ and a fixed stopping time $\theta \in \cT$, we denote by $\cE^c_0(Y_{\theta })$ the unique $y\in\rr$ such that there exists a strategy $\psi\in\Psi (y,-A)$ satisfying  $V_{\theta }(y,\psi )=Y_{\theta}$.  It is also easy to check that
\begin{equation} \label{eq2.5bb}
x_2-\cE^c_0(Y_{\tau})=\big\{p\in\rr \,|\ \exists\,\psi\in\Psi(x_2-p,-A)\!: V_{\tau}(x_2-p,\psi)=Y_{\tau}\big\}.
\end{equation}

As for the hedger (see Assumption \ref{ass1.1h}), we postulate that the game contract is replicable for the counterparty.

\begin{assumption}  
\label{ass1.1c} {\rm
We assume that the process $J^c $ is {\it replicable} for the counterparty, in the sense that for a given $x_2\in\rr$ and every $\sigma,\tau \in\cT$ there exists a pair $(q,\psi)\in \rr\times\Psi(x_2+q,-A)$ such that $V_{\theta}(x_2+q,\psi)=J^c (\sigma,\tau)$.}
\end{assumption}

For the counterparty, we have the following definition of the optimal replication, which corresponds to Definition \ref{def4.4xa} for the hedger.

\begin{definition} \label{def4.5xa} {\rm
We say that a triplet $(v^c_0(x_2),\sigmasc ,\tausc )\in\rr\times\cT\times\cT $ is a solution to the {\it counterparty's optimal replication problem} for the game contract $\cCg$ if $v^c_0(x_2)=x_2-\cE^c_0(J^c(\sigmasc ,\tausc ))$
where the stopping times $\sigmasc$ and $\tausc$ are such that}
\begin{equation} \label{eq4.5b}
\cE^c_0(J^c(\sigmasc ,\tausc ))=\min_{\tau\in\cT }\,\max_{\sigma\in\cT } \,\cE^c_0(J^c(\sigma ,\tau)).
\end{equation}
\end{definition}

\begin{assumption} 
\label{lem4.5bw}
{\rm The counterparty's optimal replication problem for the game contract $\cCg$ has a solution $(v^c_0(x_2),\sigmasc ,\tausc )$.
Furthermore, for $p^{*,c}=v^c_0(x_2)$ there exists $ \psi^*\in\Psi (x_2-p^{*,c},-A)$ such that the triplet
$(p^{*,c},\psi^*,\tausc)$ satisfies (SH$'$) and the quadruplet $(p^{*,s},\psi^*,\sigmasc ,\tausc)$ satisfies (BE$'$), so that $\cK^{r,c}(x_2) \ne \emptyset $.}
\end{assumption}

In view of Definition \ref{defsscx}, the triplet $(p^{*,c},\psi^*,\sigmasc)$ satisfies the inequality $V_{\sigmasc\wedge t}(x_2-p^{*,c},\psi^*)\ge J^c(\sigmasc,t)$ for all $t\in[0,T]$.  The proof of Lemma \ref{lem4.4cc} is omitted since it is analogous to the proofs of Lemmas \ref{lem4.4d} and \ref{lem4.4c}.

\begin{lemma} \label{lem4.4cc}
If Assumptions \ref{ass1.1s}--\ref{ass1.1bw} and \ref{ass1.1c}--\ref{lem4.5bw} are met, then we have $\psup^{s,c}(x_2) \leq v^c_{0}(x_2)$ and $\pinf^{f,c}(x_2)\geq v^c_{0}(x_2)$.
\end{lemma}

We are in a position to state the counterparty's version of Theorem \ref{the4.4}. The proof of Theorem \ref{the4.5} is based on similar arguments as the proof of Theorem \ref{the4.4} and thus it is not presented here.

\begin{theorem} \label{the4.5}
Let Assumptions \ref{ass1.1s}--\ref{ass1.1bw} and \ref{ass1.1c}--\ref{lem4.5bw} be satisfied. If $p\in\cK^{a,c}(x_2)$, then the inequality $p<v^c_{0}(x_2)$ holds and thus $\pmin^{f,c}(x_2)=v^c_{0}(x_2)$. Moreover, $\cK^{f,r,c}(x_2) \ne \emptyset $ and
\begin{equation}  \label{eq4.5f}
v^c_{0}(x_2)=\pmin^{f,c}(x_2)=\wh{p}^{f,r,c}(x_2)=\wh{p}^{s,c}(x_2).
\end{equation}
The stopping time $\sigmasc $ is a counterparty's break-even time for the triplet $(\wh{p}^{f,r,c}(x_2),\psi^{f,r,c},\tausc)$
where a trading strategy $\psi = \psi^{f,r,c}$ is implicitly given by equation \eqref{eq2.5bb} with
$(p,\sigma,\tau) = (\wh{p}^{f,r,c}(x_2), \sigmasc, \tausc )$.
More explicitly, a trading strategy $\psi^{f,r,c}$ belongs to $\Psi ( x_2-\wh{p}^{f,r,c}(x_2),- A)$ and is such that
\[
V_{\sigmasc \wedge\tausc }\big(x_2-\wh{p}^{f,r,c}(x_2) ,\psi^{f,r,c}\big)=J^h\big(\sigmasc,\tausc\big).
\]
\end{theorem}

As for the hedger, we can now define the counterparty's acceptable price for $\cCg$.

\begin{definition} \label{def1.17c}
{\rm If the set $\cK^{f,r,c}(x_2)$ is a singleton, then its unique element is denoted as $p^c(x_2)$ and called the {\it
counterparty's acceptable price} for $\cCg$.}
\end{definition}

\begin{remark}{\rm Similarly to Remark \ref{remark for the hedger's acceptable price}, under Assumptions  \ref{ass1.1s}--\ref{ass1.1bw} and \ref{ass1.1c}--\ref{lem4.5bw}, one can show that the counterparty's acceptable price for the game contract $\cCg$ is well defined. Observe that unilateral acceptable prices for $\cCg$, as computed by the hedger and counterparty, will not coincide, in general.}
\end{remark}


\section{A BSDE Approach to Game Options}   \label{sec3}

In this section, we re-examine and extend a BSDE approach to the valuation of game options in nonlinear market, which was initiated by Dumitrescu et al.~\cite{DQS2017}. Our main goal is to show that unilateral acceptable prices for a game contract $\cC^g$ can be characterized in terms of solutions to DRBSDEs driven by a multi-dimensional continuous semimartingale $S$.  In this section, we postulate that the wealth process $V=V(y,\phi,A)$ satisfies the SDE
\begin{equation} \label{ueq3x}
V_t=y-\int_0^t g(u,V_u,\xi_u)\,du+\int_0^t\xi_u\,dS_u+A_t,
\end{equation}
where $y \in \rr$ and a process $\xi$ are given (recall also that $A_0=0$). By applying Lemma 3.1 in \cite{KNR2018} with $y_1=x+p<x+p'=y_2,\,f_1=f_2=g$ and $z=\xi$, one can check that the condition of strict monotonicity of wealth (see Assumption \ref{ass1.1s}) is met when the dynamics of the wealth process are given by \eqref{ueq3x}, for instance, the mapping $g=g(t,v,z)$ is Lipschitz continuous with respect to $v$.

\subsection{Dynamics of the Wealth Process} \label{sec3.1}

We will briefly describe the main features of the mechanism of nonlinear trading, which generates the wealth dynamics given by \eqref{ueq3x}. We first introduce the notation for traded assets in our market model, which are: cash accounts, risky assets, and funding accounts associated with risky assets. It should be stressed that results in this section do not depend on the choice of a particular model for primary assets and trading arrangements.

Let $\cS=(S^1,\ldots,S^d)$ stand for the collection of prices of a family of $d$ risky assets where the processes $S^1,\dots,S^d$ are continuous semimartingales. Continuous processes of finite variation, denoted as $B^{0,l}$ and $B^{0,b}$, represent the {\it lending} {\it borrowing} unsecured cash accounts, respectively. For every $j=1,2,\ldots,d$, we denote by $B^{j,l}$ (respectively, $B^{j,b}$) the {\it lending} (respectively, {\it borrowing}) {\it funding account} associated with the $i$th risky asset, and also assumed to be continuous processes of finite variation. The financial interpretation of these accounts varies from case to case (for more details, see \cite{BCR2018,BR2015}).  Let us denote by $\cB$ the collection of all cash and funding accounts available to a trader. For simplicity of presentation, we maintain our assumption that the hedger and the counterparty have identical market conditions but it is clear that this assumption is not relevant for our further developments and thus it can be easily relaxed. A {\it trading strategy} is an ${\mathbb R}^{3d+2}$-valued, $\gg$-adapted process $\phi=( \xi^1,\ldots,\xi^{d}; \psi^{0,l},\psi^{0,b},\dots ,\psi^{d,l}, \psi^{d,b})$ where the components represent all outstanding positions in the risky assets $S^j,\, j=1,2, \dots,d$, cash accounts $B^{0,l},\, B^{0,b}$, and funding accounts $B^{j,l}, B^{j,b} ,\, j=1,2, \dots,d$ for risky assets.

\begin{definition}\label{self financing} {\rm
We say trading strategy $(y,\phi)$ is {\it self-financing} for $\cC^g$ and we write $\phi \in \Psi (y,A)$ if the {\it wealth process} $V(y,\phi,A)$, which is given by
\begin{equation*} 
V_t(y,\phi,A)=\sum_{j=1}^{d}\xi^j_t S^j_t+\sum_{j=0}^{d}\psi^{j,l}_t B^{j,l}_t+\sum_{j=0}^{d}\psi^{j,b}_t B^{j,b}_t ,
\end{equation*}
satisfies, for every $t \in [0,T]$,
\begin{equation*} 
V_t(y,\phi,A)=y+\sum_{j=1}^{d}\int_0^t\xi^j_u\,dS^j_u+\sum_{j=0}^{d}\int_0^t\psi^{j,l}_u\,dB^{j,l}_u+\sum_{j=0}^{d}\int_0^t\psi^{j,b}_u\,dB^{j,b}_u + A_t
\end{equation*}
subject to additional constraints imposed on the components of $\phi $. In particular, we postulate that $\psi^{j,l}_t \geq 0,\, \psi^{j,b}_t \leq 0$ and $\psi^{j,l}_t \psi^{j,b}_t=0$ for all $j=0,1, \dots ,d$ and $t \in [0,T]$.
} \end{definition}

Due to additional trading constraints, which depend on the particular trading mechanism, the choice of an initial value $y$ and a process $\xi $ is known to uniquely specify the wealth process of a self-financing strategy $\phi\in\Psi (y,A)$. In addition, one needs also to introduce some form of {\it admissibility} of trading strategies and to postulate that the market model $\cM=(\cB ,\cS ,\Psi (A))$ where the class $\Psi (A) = \cup_{y\in \rr}\Psi (y,A)$ of admissible trading strategies is arbitrage-free in a suitable sense, for instance, the market model $\cM$ can be assumed to be {\it regular}, in the sense of Bielecki et al.~\cite{BCR2018}. It is important to notice that, due to the trading constraints, differential funding costs and possibly also some additional adjustment processes, which are not explicitly stated in Definition \ref{self financing}, the dynamics of the wealth process are nonlinear, in general. We refer the reader to Bielecki et al.~\cite{BCR2018,BR2015} for more details on the self-financing property of a nonlinear trading strategy and to Nie and Rutkowski~\cite{NR2,NR3,NR4} for explicit examples of nonlinear markets.

%

\subsection{Comparison Properties of Nonlinear Evaluations} \label{sec3.2}

Our goal in this section is to examine the valuation and hedging of game contracts for the special case where the wealth dynamics are driven by \eqref{ueq3x}. Therefore, we henceforth postulate that the wealth process $V=V(y,\phi,A)$ is governed by the SDE \eqref{ueq3x} where $\xi=(\xi^1, \dots,\xi^d)$ is given and the mapping $g$ satisfies some additional assumptions.
To keep the presentation succinct and covering several alternative nonlinear market models, we will directly postulate that the associated (possibly doubly reflected) BSDEs enjoy desirable properties, such as: the existence, uniqueness, and the strict comparison property for solutions to BSDEs, which are known to hold under various circumstances. As was already mentioned, a particular instance of a market model given by \eqref{ueq3x} has been studied in detail by Dumitrescu et al.~\cite{DQS2017}.

We will use the standard terminology related to nonlinear evaluations generated by solutions to BSDEs (see, e.g., Chapter 3 in Peng~\cite{P2004a} or Section 4 in Peng \cite{P2004b}). Consider the following BSDE on $[0,s]$
\begin{equation}\label{xBSDEU}
Y_t =\zeta_s +\int_t^s g(u,Y_u,Z_u)\,du-\int_t^s Z_u\,dM_u-(\hHh_s-\hHh_t),
\end{equation}
where $\zeta_s \in L^2(\cG_s)$, $M$ is a $d$-dimensional martingale, the process $\hHh$ is real-valued and $\gg$-adapted,
and the generator $g: \Omega\times[0,T]\times\rr\times\rr^d \rightarrow \rr$ is $\cP \otimes \cB (\rr )\otimes\cB (\rr^d)/\cB(\rr)$-measurable where $\cP$ is the $\sigma$-field of predictable sets on $\Omega\times[0,T]$. Assume that the BSDE \eqref{xBSDEU} has a unique solution $(Y,Z)$ in a suitable space of stochastic processes (see, e.g., \cite{CFS2008,NR2}).
For every $0\leq t\leq s \leq T$ and $\zeta_s \in L^2(\cG_s)$, we denote $\cEgH_{t,s}(\zeta_s )= Y_t$ where $(Y,Z)$ solves the BSDE \eqref{xBSDEU} with $Y_s=\zeta_s$. Then the system of operators $\cEgH_{t,s}: L^2(\cG_s) \to L^2(\cG_t)$ is called the {\it $\cEgH$-evaluation}.  It is worth noting that a deterministic dates $t \leq s$ appearing in the BSDE \eqref{xBSDEU} can be replaced by arbitrary $\gg$-stopping times $\tau \leq \sigma $ from $\cT$ and thus the notion of the $\cEgH$-evaluation can be extended to stopping times $\cEgH_{\tau ,\sigma }: L^2(\cG_\sigma ) \to L^2(\cG_\tau )$. The concept of the (strict) comparison property is of great importance in the theory of BSDEs and nonlinear evaluations.

\begin{definition} \label{xdefmong} {\rm
We say that the {\it comparison property} of $\cEgH$ holds if for every stopping time $\tau\in\cT$ and random variables $\zeta^1_{\tau},\zeta^2_{\tau} \in L^2(\cG_{\tau})$, the following property is valid: if $\zeta^1_{\tau} \geq \zeta^2_{\tau}$ then $\cEgH_{0,\tau }(\zeta^1_{\tau})\geq \cEgH_{0,\tau }(\zeta^2_{\tau})$. We say that the {\it strict comparison property} of $\cEgH$ holds if for every $\tau\in\cT$ and $\zeta^1_{\tau},\zeta^2_{\tau} \in L^2(\cG_{\tau})$ if $\zeta^1_{\tau}\geq\zeta^2_{\tau}$ and $\zeta^1_{\tau}\ne\zeta^2_{\tau}$ then $\cEgH_{0,\tau}(\zeta^1_{\tau})>\cEgH_{0,\tau}(\zeta^2_{\tau})$.}
\end{definition}

The nonlinear evaluation given by solutions to the BSDE \label{xnBSDEU} with $M=S$ and $\hHh =A$ is interpreted as the hedger's nonlinear evaluation. Note that here $S$ is the process of (possibly discounted) assets prices and thus it is assumed to be
a $d$-dimensional continuous martingale after a change of a probability measure.

\begin{definition} \label{defcEg}
{\rm The nonlinear evaluation $\cE^{g,A}$ associated with the BSDE
\begin{equation} \label{nhBSDE}
Y_t=\zeta_s+\int_t^s g(u,Y_u,Z_u)\,du-\int_t^s Z_u\,dS_u-(A_s-A_t)
\end{equation}
is denoted by $\cEgh$ and called the {\it hedger's $g$-evaluation} for $A$.}
\end{definition}

In Section \ref{sec3.4} and \ref{sec3.4b}, we address the hedger's pricing problem and we work under the following standing assumption.

\begin{assumption} \label{assnBSDE} {\rm We postulate that:\\
(i) the wealth process $V=V(y,\phi,A)$ of any trading strategy $\phi \in \Psi (y,A)$ satisfies \eqref{ueq3x},\\
(ii) for any fixed $y \in \rr$ and any process $\xi$ such that the stochastic integral in \eqref{ueq3x} is well defined,
the SDE \eqref{ueq3x} has a unique strong solution, \\
(iii) the strict monotonicity property holds for the wealth $V(y,\phi,A)$ (see Assumption \ref{ass1.1s}),\\
(iv) for every $(s,\zeta_s)\in [0,T]\times L^2(\cG_s)$ the BSDE \eqref{nhBSDE} has a unique solution $(Y,Z)$ on $[0,s]$, \\
(v) the inequalities $X^h_t<X^c_t$ and $X^h_t \leq \Xxb_t \leq X^c_t$ hold for all $t \in [0,T]$.}
\end{assumption}

In Sections \ref{sec3.5} and \ref{sec3.5b}, when studying the counterparty's unilateral valuation and exercising problems, we will use a slight modification of Assumption \ref{assnBSDE} with the process $A$ replaced by $-A$ and with suitably modified assumptions about BSDEs and the counterparty's $g$-evaluation $\cEgc$.

\begin{remark} {\rm
In view of Lemma 3.1 in Kim et al. \cite{KNR2018} and condition (ii) in Assumption \ref{assnBSDE}, condition (iii) in Assumption \ref{assnBSDE} is not restrictive, since it is valid for every generator $g$. For explicit assumptions about the generator $g$ ensuring that the BSDE \eqref{nhBSDE}, where $S$ is a multi-dimensional, continuous, square-integrable martingale enjoying the predictable representation property, has a unique solution and the strict comparison property of the  hedger's $g$-evaluation $\cEgh$ holds, the reader is referred to Theorems 3.2 and 3.3 in Nie and Rutkowski~\cite{NR2}.}
\end{remark}

\subsection{Doubly Reflected BSDEs} \label{sec3.3}

We first introduce the general notation for a doubly reflected BSDE (DRBSDE).
Let $\zeta^l$ and $\zeta^u$ be the two $\gg$-adapted, c\`adl\`ag processes such that $\zeta^l_t <\zeta^u_t$ for all $t \in [0,T]$. Also, let $\zeta^m_T \in L^1(\cG_T)$ be a random variable such that $\zeta^l_T \leq \zeta^m_T \leq \zeta^u_T$. We work under Assumption \ref{assnBSDE} regarding the dynamics of the wealth process and the properties of solutions to BSDE $(g,\zeta^m_T )$.
In addition, we will make additional assumptions about the existence and uniqueness of solutions to a DRBSDE for the hedger and counterparty. They can be justified by results established in numerous papers; see, for instance, Bayraktar and Yao~\cite{BY2015}, Cvitani\'c and Karatzas~\cite{CK1996}, Cr\'epey and Matoussi~\cite{CM2008}, Essaky and Hassani~\cite{EH2013}, Dumitrescu et al.~\cite{DQS2016}, Grigorova et al.~\cite{GIOQ2017}, Hamad\`ene and Ouknine~\cite{HO2016}, Klimsiak \cite{K2013, K2015} or Lepeltier and Xu \cite{LX2007} and the references therein. In particular, Assumption \ref{assRBSDEh} is known to be satisfied by several stochastic market models that were studied in papers on the so-called valuation adjustments.

The definition of a solution to the DRBSDE with obstacles $\zeta^l$ and $\zeta^u$, as stated in Assumption  \ref{assRBSDEh}, is a minor modification of Definition 2.4 in Dumitrescu et al.~\cite{DQS2016}. For alternative versions of the minimality (Skorokhod) conditions in the definition of a solution to DRBSDE with progressively measurable (or $\gg$-optional) obstacles, we refer to Klimsiak \cite{K2013, K2015} and Grigorova et al.~\cite{GIOQ2017}. Note that we deliberately do not specify particular spaces
of stochastic processes in which the components $Y$ and $Z$ are searched for, since our further results do not depend on
the choice of these spaces. Only the properties of the processes $\Lll$ and $\Uuu$ in a unique solution $(Y,Z,\Lll,\Uuu)$ to the hedger's DRBSDE \eqref{hDRBSDE} (and, analogously, of the processes $\cLll$ and $\cUuu$ in a unique solution $(y,z,\cLll,\cUuu)$ to the counterparty's RBSDE  \eqref{cDRBSDE}) will play an essential role and thus they are stated explicitly and analyzed in detail.

\begin{assumption} \label{assRBSDEh}
{\rm The DRBSDE with parameters $(g,\zeta^l,\zeta^u,\zeta^m_T)$
\begin{equation} \label{DRBSDEg}
\left\{ \begin{array} [c]{ll}
-dY_t=g(t,Y_t,Z_t)\,dt-Z_t\,dS_t-dA_t+d\Lll_t-d\Uuu_t,\quad Y_T=\zeta^m_T ,\medskip\\
\zeta^l_t\leq Y_t\leq \zeta^u_t,\quad \int_0^T (Y_t-\zeta^l_t)\,d\Lll_t^c=\int_0^T (\zeta^u_t-Y_t)\,d\Uuu_t^c=0,
\medskip\\
\Delta \Lld=-\Delta (Y-A) \I_{\{ Y_{-}= \zeta^l_{-}\}},\quad \Delta \Uud=\Delta (Y-A) \I_{\{ Y_{-}= \zeta^u_{-}\}},
\end{array} \right.
\end{equation}
has a unique solution $(Y,Z,\Lll,\Uuu)$ where the processes $\Lll,\Uuu$ are $\gg$-predictable, c\`adl\`ag, nondecreasing and such that $\Lll_0=\Uuu_0=0$. Moreover, the equalities $\Lll=\Llc+\Lld$ and $\Uuu=\Uuc+\Uud$ give their unique decompositions into continuous and jump components.}
\end{assumption}

Dumitrescu et al.~\cite{DQS2016} provide a thorough examination of a particular instance of a DRBSDE driven by a Brownian motion and a compensated random measure. They postulate, using the present setup and notation, that the conditions $\Delta \Lld_\tau=-\Delta (Y_\tau-A_\tau) \I_{\{ Y_{\tau-}= \zeta^l_{\tau-}\}}$ and $\Delta \Uud_\tau=\Delta (Y_\tau-A_\tau) \I_{\{ Y_{\tau-}= \zeta^u_{\tau-}\}}$ are satisfied for every $\gg$-predictable stopping time $\tau $.  Since $\Lll$ and $\Uuu $ are $\gg$-predictable, nondecreasing processes, these conditions are equivalent to
$\Delta \Lld=-\Delta (Y-A) \I_{\{ Y_{-}= \zeta^l_{-}\}}$ and $\Delta \Uud=\Delta (Y-A) \I_{\{ Y_{-}= \zeta^u_{-}\}}$ where the
equality means that the processes in the right- and left-hand sides are indistinguishable.

\begin{remark} \label{remcon} {\rm
Using the arguments from the proof of Theorem 3.7 of Dumitrescu et al.~\cite{DQS2016}, it is easy to show that if the process $\zeta^l-A$ (respectively, $-\zeta^u+A$) is left-upper-semicontinuous, then the process $\Lll$ (respectively, $\Uuu$) is continuous. This important observation motivates Assumption \ref{assRBSDEz} for the hedger, as well as the analogous Assumption \ref{assRBSDEcz} for the counterparty.}
\end{remark}

Let $\zeta^m$ be a $\gg$-adapted, c\`adl\`ag process such that $\zeta^l_t\leq \zeta^m_t\leq \zeta^u_t$ for all $t \in [0,T]$
and let the payoff $J(\zeta^l,\zeta^u,\zeta^m ,\sigma,\tau)$ of a stochastic stopping game be given by
\begin{equation} \label{dynpay}
J(\zeta^l,\zeta^u,\zeta^m ,\sigma,\tau):=\zeta^l_{\tau}\I_{\{\tau<\sigma\}}+\zeta^u_\sigma\I_{\{\sigma<\tau\}}+\zeta^m_{\tau}\I_{\{\tau=\sigma\}}.
\end{equation}
Then the upper and lower values of the nonlinear Dynkin game associated with the hedger's $g$-evaluation $\cEgh$ and with
the payoff $J(\zeta^l,\zeta^u,\zeta^m ,\sigma,\tau)$ are defined by the following expressions
\begin{align*}
&\overline{\cV}^h_0(\zeta^l,\zeta^u,\zeta^m):=\inf_{\sigma\in\cT}\sup_{\tau\in\cT}\cEgh_{0,\sigma\wedge\tau}\big(J(\zeta^l,\zeta^u,\zeta^m,\sigma,\tau)\big),\\
&\underline{\cV}^h_0(\zeta^l,\zeta^u,\zeta^m):=\sup_{\tau\in\cT}\inf_{\sigma\in\cT}
\cEgh_{0,\sigma\wedge\tau}\big(J(\zeta^l,\zeta^u,\zeta^m ,\sigma,\tau)\big),
\end{align*}
so that, manifestly, the inequality $\overline{\cV}^h_0(\zeta^l,\zeta^u,\zeta^m)\geq \underline{\cV}^h_0(\zeta^l,\zeta^u,\zeta^m)$ is always true.

In view of the existing results on linear and nonlinear Dynkin games (see, in particular, Theorems 3.5, 3.7, 4.8 and 4.10 in Dumitrescu et al.~\cite{DQS2016})),  it would be tempting to postulate that the above-mentioned nonlinear Dynkin game has the {\it value}, in the sense that the equality $\overline{\cV}^h_0(\zeta^l,\zeta^u,\zeta^m)=\underline{\cV}^h_0(\zeta^l,\zeta^u,\zeta^m)$ holds and, moreover, that the value of the  game coincides with the initial value $Y_0$ of a solution to the DRBSDE \eqref{DRBSDEg}, that is, the equality $Y_0=V^h_0(\zeta^l,\zeta^u,\zeta^m)$ holds where
\[
V^h_0(\zeta^l,\zeta^u,\zeta^m):=\overline{\cV}^h_0(\zeta^l,\zeta^u,\zeta^m)=\underline{\cV}^h_0(\zeta^l,\zeta^u,\zeta^m).
\]
This relationship between the nonlinear Dynkin game with the payoff given by \eqref{dynpay} and a solution to the DRBSDE \eqref{DRBSDEg}  is known to be satisfied in some setups and thus it could be postulated as well. However, as we will argue in what follows, in fact it would not be natural to postulate in the present context that the nonlinear Dynkin game has a value, since its existence is completely immaterial for unilateral valuation, hedging and exercising results. In fact, it suffices to either to assume or to demonstrate for each particular instance of a nonlinear market model under study that the equality $Y_0=\inf_{\sigma\in\cT}\sup_{\tau\in\cT}\cEgh_{0,\sigma\wedge\tau} (J(\zeta^l,\zeta^u,\zeta^m ,\sigma,\tau))$ is valid for
the hedger's and counterparty's nonlinear Dynkin games.

Put another way, it is not necessary to postulate that the nonlinear Dynkin game associated with the game contract has the value, since only the upper value $\overline{\cV}^h_0(X^l,X^u,X^m)$ matters for the hedger. Similarly, the counterparty's price will be expressed in terms of the upper value $\overline{\cV}^c_0(x^l,x^u,x^m)$ where, in general, the processes $x^l,x^u$ and $x^m$ do not coincide with $X^l,X^u$ and $X^m$, respectively.

We conclude that it order to cover both the case of the hedger and that of the counterparty, it is convenient to introduce the following assumption, which is supported by results in Dumitrescu et al.~\cite{DQS2016} and Grigorova et al.~\cite{GIOQ2017}.

\begin{assumption} \label{assDGw}
{\rm Let $\zeta^m$ be a $\gg$-adapted, c\`adl\`ag process such that $\zeta^l_t\leq \zeta^m_t\leq \zeta^u_t$ for all $t \in [0,T]$. We postulate that the following equalities hold
\[
Y_0=\overline{\cV}^h_0(\zeta^l,\zeta^u,\zeta^m):=\inf_{\sigma\in\cT}\sup_{\tau\in\cT}\cEgh_{0,\sigma \wedge \tau}
\big(J(\zeta^l,\zeta^u,\zeta^m,\sigma,\tau)\big)=\cEgh_{0,\sigma^*\wedge \tau^*}\big(J(\zeta^l,\zeta^u,\zeta^m,\sigma^*,\tau^*)\big)
\]
where $\sigma^*:=\inf\{t\in [0,T]\,|\,Y_t=\zeta^u_t\}$ and $ \tau^*:=\inf\{t\in [0,T]\,|\,Y_t=\zeta^l_t\}$.}
\end{assumption}

\begin{remark} \label{rem5.1} {\rm
In the special case when $S=W$ is a Brownian motion, one may also refer to Bayraktar and Yao~\cite{BY2015} (see Assumptions (H.1)--(H.5) and Theorem 2.1 in \cite{BY2015}) for explicit assumptions ensuring that a particular nonlinear Dynkin game has the value, a unique solution $(Y,Z,\Lll,\Uuu)$ exists in a suitable space of stochastic processes, and the equality $Y_0=\overline{\cV}^h_0(\zeta^l,\zeta^u,\zeta^m)$ holds (in fact, they show that $Y_0=\cV^h_0(\zeta^l,\zeta^u,\zeta^m)$).}
\end{remark}

\subsection{Hedger's Acceptable Price via DRBSDE} \label{sec3.4}

In Sections \ref{sec3.4} and \ref{sec3.4b},
it is postulated throughout that Assumptions \ref{assnBSDE}--\ref{assDGw} are satisfied by the hedger's wealth and the hedger's DRBSDE with parameters $(g,X^l,X^u,X^m_T)$ where the processes $X^l,X^u$ and the random variable $X^m_T$ are given in the statement of Proposition \ref{pro5.1} (see also Section \ref{sec2.4}). Note that the obstacles $X^l$ and $X^u$ are c\`adl\`ag processes, since the processes $X^h,X^c$ and $\Vben(x_1)$ are assumed to be c\`adl\`ag. Finally, recall that the hedger's relative reward $J^h(\sigma,\tau)=J(x_1,X^l,X^u,X^m,\sigma,\tau)$ is given by (see equation \eqref{eq4.4h})
\begin{equation} \label{eq.5.0}
J^h(\sigma,\tau):=\Vben_{\sigma\wedge\tau}(x_1)-I(\sigma,\tau)
=X^l_{\tau}\I_{\{\tau<\sigma\}}+X^u_\sigma\I_{\{\sigma<\tau\}}+X^m_{\sigma}\I_{\{\tau=\sigma\}}.
\end{equation}
We start by analyzing the lower bound for the hedger's superhedging costs.
The following result corresponds to Proposition 3.4 in Dumitrescu et al. \cite{DQS2017} where, however, the nonlinear Dynkin
game associated with a particular game option is also shown to have the value.

\begin{proposition}  \label{pro5.1}
If 
the comparison property of the hedger's $g$-evaluation holds and the process $-X^u+A$ is left-upper-semicontinuous, then the lower bound for the hedger's superhedging costs satisfies
\begin{equation} \label{eq.5.1}
\pinf^{s,h}(x_1)=Y_0-x_1=\inf_{\sigma\in\cT}\sup_{\tau\in\cT}\cEgh_{0,\sigma\wedge\tau}(J^h(\sigma,\tau))-x_1,
\end{equation}
where $(Y, Z, \Lll, \Uuu)$ is the unique solution to the hedger's DRBSDE with parameters $(g,X^l,X^u,X^m_T)$
\begin{equation} \label{hDRBSDE}
\left\{ \begin{array} [c]{ll}
-dY_t=g(t,Y_t,Z_t)\,dt-Z_t\,dS_t-dA_t+d\Lll_t-d\Uuu_t,\quad Y_T=X^m_T ,\medskip\\
X^l_t\leq Y_t\leq X^u_t,\quad \int_0^T (Y_t-X^l_t)\,dL_t^c=\int_0^T (X^u_t-Y_t)\,dU_t^c=0,\medskip\\
\Delta \Lld=-\Delta (Y-A) \I_{\{ Y_{-}=X^l_{-}\}},\quad \Delta \Uud=\Delta (Y-A) \I_{\{ Y_{-}= X^u_{-}\}},
\end{array} \right.
\end{equation}
where $X^l_t:=\Vben_t(x_1)-X^c_t<X^u_t:=\Vben_t(x_1)-X^h_t$ for all $t \in [0,T]$ and $X^l_T\leq X^m_T:=\Vben_T(x_1)-\Xxb_T\leq X^u_T$.
\end{proposition}

\proof
We fix $x_1\in\rr$ and we first prove that $\pinf^{s,h} (x_1) \le Y_0-x_1$. It suffices to show that for the initial value $p'=Y_0-x_1$, where $Y_0$ is obtained from the DRBSDE (\ref{hDRBSDE}), we can find a hedger's superhedging strategy. Our goal is thus to show that there exists a strategy $(\phi',\sigma')\in\Psi(x_1 + p',A)\times\cT$ such that the triplet $(p',\phi',\sigma')$ fulfills condition (SH). To this end, we define the strategy $(p',\phi')=(Y_0-x_1,Z)$ where $(Y, Z, \Lll, \Uuu)$ is a unique solution to the DRBSDE (\ref{hDRBSDE}).  Then, for a given process $Z$, the wealth $V=V(x_1 + p',\phi')$ satisfies the forward SDE
\begin{equation} \label{eq.fSDE2}
\left\{ \begin{array} [c]{ll}
dV_t=-g(t,V_t,Z_t)\,dt+Z_t\,dS_t+dA_t ,\medskip\\
V_0=Y_0.
\end{array} \right.
\end{equation}
Furthermore, if we set $\sigma^h:=\inf \{t\in [0,T]\,|\,Y_t=X^u_t\}$, then using the continuity of $U$ (see Remark \ref{remcon}),  we obtain $\Uuu_{\sigma^h}=0$, and thus, $\Uuu=0$ on the stochastic interval $[0,\sigma^h]$ , the DRBSDE (\ref{hDRBSDE}) can be viewed as a forward SDE for $Y$ with given processes $Z$ and $\Lll$
\begin{equation} \label{eq.fbSDE2}
\left\{ \begin{array} [c]{ll}
dY_t=-g(t,Y_t,Z_t)\,dt+Z_t\,dS_t+dA_t-d\Lll_t ,\medskip\\
Y_0=Y_0,\quad X^l_t \leq Y_t \leq X^u_t,\quad \int_0^T (Y_t-X^l_t)\,dL_t^c=0,\quad \Delta \Lld=-\Delta (Y-A) \I_{\{ Y_{-}=X^l_{-}\}}.
\end{array} \right.
\end{equation}
From Lemma 3.1 in \cite{KNR2018}, we infer that $V_t \ge Y_t$ for all $t \in [0,\sigma^h]$ and thus, since $Y_t \ge X^l_t$ for all $t \in [0,T]$, we conclude that $V_t \ge X^l_t$ for all $t \in [0,\sigma^h]$.  Moreover, $V_{\sigma^h} \ge Y_{\sigma^h}
=X^u_{\sigma^h} \geq X^m_{\sigma^h}$.  Consequently, for all $t\in [0,T]$,
\[
V_{\sigma^h \wedge t} \ge X^l_t\I_{\{t< \sigma^h\}}+X^u_{\sigma^h}\I_{\{\sigma^h <t\}}+X^m_t\I_{\{t=\sigma^h\}}
=J^h(\sigma^h,t),
\]
which means that $(p',\phi',\sigma')=(Y_0-x_1,Z,\sigma^h)$ is a hedger's superhedging strategy, in the sense of Definition \ref{defssxx}. We conclude that the inequality $\pinf^{s,h}(x_1) \le Y_0-x_1$ is valid.

It now remains to show that $\pinf^{s,h} (x_1) \ge Y_0-x_1$.  By the definition of a supremum, it suffices to show that $Y_0-x_1$ is a lower bound for any $p\in\cK^{s,h}(x_1)$. Let us take an arbitrary $p\in\cK^{s,h}(x_1)$. Then there exists a pair $(\phi,\sigma)\in\Psi(x_1+p,A)\times\cT$ such that, for all $\tau\in\cT$,
\[
V_{\sigma \wedge \tau}(x_1+p,\phi)\ge X^l_\tau \I_{\{\tau\le\sigma\}}+X^u_{\sigma}\I_{\{\tau>\sigma\}}+X^m_{\sigma}\I_{\{\tau=\sigma\}}=J^h(\sigma,\tau).
\]
The comparison property of the hedger's $g$-evaluation yields, for every $\tau\in\cT$,
\[
x_1+p=\cEgh_{0,\sigma \wedge \tau}\big(V_{\sigma \wedge \tau}(x_1+p,\phi)\big) \ge \cEgh_{0,\sigma \wedge \tau} (J^h(\sigma,\tau)).
\]
Since $\tau\in\cT$ is arbitrary, we obtain
\[
x_1+p\ge\sup_{\tau\in\cT}\cEgh_{0,\sigma\wedge\tau}(J^h(\sigma,\tau))\ge\inf_{\sigma\in\cT}\sup_{\tau\in\cT}
\cEgh_{0,\sigma\wedge\tau}(J^h(\sigma,\tau))=Y_0 ,
\]
where the last equality follows from Assumption \ref{assDGw}. We conclude that $\pinf^{s,h}(x_1)\ge Y_0-x_1$,
which completes the proof of equality \eqref{eq.5.1}.
\finproof

For the justification of the next assumption, we refer to Remark \ref{remcon}.

\begin{assumption} \label{assRBSDEz}
{\rm  The processes $X^l-A$ and $-X^u+A$ are left-upper-semicontinuous so that the processes $\Lll$ and $\Uuu$
in the solution $(Y, Z, \Lll, \Uuu)$ to the hedger's DRBSDE \eqref{hDRBSDE} are continuous.}
\end{assumption}

In the next result, we identify a hedger's replicating strategy for $\cCg$.  Recall that the concept of a hedger's replicating strategy for $\cCg$ was introduced in Section \ref{sec2.3}. Note that when defining stopping times, we henceforth adopt the convention that $\inf \emptyset = T$.

\begin{proposition}  \label{pro5.2}
If Assumption \ref{assRBSDEz} is satisfied and the comparison property of the hedger's $g$-evaluation holds, then the following assertions are valid: \\
(i) the triplet $(Y_0-x_1,Z,\sigma^h)$ is a hedger's replicating strategy for $\cCg$, where $(Y,Z,\Lll,\Uuu)$ is the unique solution to the DRBSDE (\ref{hDRBSDE}) and the stopping time $\sigma^h:=\inf\{t\in [0,T]\,|\,Y_t=X^u_t\}$,\hfill \break
(ii) the hedger's minimum superhedging and replication costs satisfy
\begin{equation}  \label{eq3.3b}
\pmin^{r,h}(x_1)=\pmin^{s,h}(x_1)=Y_0-x_1,
\end{equation}
(iii) we have
\[
\pmin^{r,h}(x_1)=\cEgh_{0,\sigma^h\wedge\tau^h }(J^h(\sigma^h,\tau^h))-x_1 ,
\]
where $\tau^h:=\inf\{t \in [0,T]\,|\,Y_t=X^l_t\}.$
\end{proposition}

\proof
We already know that $Y_0-x_1=\pinf^{s,h}(x_1) \le \pinf^{r,h}(x_1)$ and thus to prove parts (i) and (ii), it suffices to show that $(p',\phi',\sigma')=(Y_0-x_1, Z,\sigma^h)$ is a hedger's replicating strategy for $\cCg$.  Noticing that the right-continuity of the processes $Y$ and $X^u$ gives $Y_{\sigma^h}= X^u_{\sigma^h}$ and the continuity of $\Lll$ and $\Uuu$ ensures that $\Lll_{\tau^h}=0$ and $\Uuu_{\sigma^h}=0$,  hence the DRBSDE (\ref{hDRBSDE}) can be viewed on $[0,\sigma^h \wedge \tau^h]$ as the SDE (\ref{eq.fbSDE2}) with a predetermined process $Z$
and the process $L$ satisfying $\Lll_t=0$ for all $t \in [0,\tau^h]$. Consequently, from the uniqueness of a solution to the SDE (\ref{eq.fbSDE2}), we obtain the equality $V_t(Y_0,Z)=Y_t$ on $[0,\sigma^h \wedge \tau^h]$. In particular, $\P (\tau^h=\sigma^h)=0$ (since $X^l<X^u$) and
\[
V_{\sigma^h\wedge\tau^h}(Y_0,Z)=Y_{\sigma^h\wedge\tau^h}=X^l_{\tau^h}\I_{\{\tau^h<\sigma^h\}}+X^u_{\sigma^h}\I_{\{\tau^h>\sigma^h\}}
=J^h(\sigma^h,\tau^h),
\]
which means that $(Y_0-x_1,Z,\sigma^h)$ is a hedger's replicating strategy for $\cCg$. Assertion (iii) is now an immediate consequence of Assumption \ref{assDGw} applied to the hedger's DRBSDE (\ref{hDRBSDE}).
\finproof

The last step in solving the hedger's valuation hinges on showing that the hedger's minimum replication cost is also his maximum fair price and to give alternative representations for the hedger's acceptable price. The proof of the next result is fairly similar to the proof of Theorem \ref{the4.4}; it is nevertheless given here for the sake of completeness.

\begin{theorem}  \label{the5.1}
If Assumption \ref{assRBSDEz} is satisfied and the strict comparison property of the hedger's $g$-evaluation holds, then the hedger's acceptable price $p^h(x_1)$ satisfies
\[
p^h(x_1)=\pmin^{f,r,h}(x_1)=\wh{p}^{f,h}(x_1)=\pmin^{r,h}(x_1)=Y_0-x_1=\overline{\cV}^h_0(X^l,X^u,X^m)-x_1.
\]
\end{theorem}

\proof
It suffices to show that $\pmin^{r,h}(x_1)$ belongs to the set $\cK^{f,h}(x_1)$ or, equivalently, that the inequality $\pmin^{r,h}(x_1) < p$ holds for every $p\in\cK^{a,h}(x_1)$.  To this end, we will argue by contradiction.  Let us denote $p'=\pmin^{r,h}(x_1)$.  Assume that $p'\in\cK^{a,h}(x_1)$ so that there exists a strategy $(\phi',\sigma')\in\Psi(x_1 + p',A)\times\cT$ such that the triplet $(p',\phi',\sigma')$ fulfills condition (AO). Then we have that, for every $\tau\in\cT$,
\[
\P\big(V_{\sigma'\wedge\tau}(x_1+p',\phi')\ge J^h(\sigma',\tau)\big)=1,\quad \P\big(V_{\sigma'\wedge\tau}(x_1+p',\phi')>J^h(\sigma',\tau)\big)>0.
\]
Setting $\tau=\tau^h$ and using the strict comparison property of the hedger's $g$-evaluation, we obtain
\[
x_1+p'=\cEgh_{0,\sigma'\wedge\tau^h}\big(V_{\sigma'\wedge\tau^h}(x_1+p',\phi')\big)>\cEgh_{0,\sigma'\wedge \tau^h}(J^h(\sigma',\tau^h)).
\]
However, in view of Assumption \ref{assDGw} and Proposition \ref{pro5.2} (iii), we have that
\[
\cEgh_{0,\sigma'\wedge \tau^h}(J^h(\sigma',\tau^h))\ge\cEgh_{0,\sigma^h\wedge\tau^h}(J^h(\sigma^h,\tau^h))=x_1+p'
\]
and thus we obtain
\[
x_1+p'=\cEgh_{0,\sigma'\wedge\tau^h}\big(V_{\sigma'\wedge\tau^h}(x_1+p',\phi')\big)>
\cEgh_{0,\sigma'\wedge\tau^h}(J^h(\sigma',\tau^h))\ge\cEgh_{0,\sigma^h\wedge\tau^h}(J^h(\sigma^h,\tau^h))=x_1+p',
\]
which is a clear contradiction.  Consequently, $\wh {p}^{r,h}(x_1) \notin \cK^{a,h}(x_1)$ and thus $\wh {p}^{r,h}(x_1)\in\cK^{f,h}(x_1)$, which means that the set $\cK^{f,r,h}(x_1)$ is nonempty.  In view of Lemma \ref{lem.price}, we conclude that $\wh{p}^{f,h}(x_1)=\pmin^{r,h}(x_1)=\pmin^{f,r,h}(x_1)=\pmin^{s,h}(x_1)$. The assertion now follows from Propositions \ref{pro5.1} and \ref{pro5.2}.
\finproof

\subsection{Hedger's Rational Cancellation Times} \label{sec3.4b}

Our next goal is to provide some useful characterizations of the class all possible hedger's rational cancellation times for the game contract $\cCg$.

\begin{definition}  \label{ratioca}
{\rm We say that $\sigma'\in\cT $ is a {\it hedger's rational cancellation time} for $\cCg$ if the contract is traded at the hedger's acceptable price $p^h(x_1)$ at time $0$ and there exists a trading strategy $\phi\in\Psi(x_1+p^h(x_1),A)$ such that $V_{\sigma' \wedge \tau}(x_1+p^h(x_1),\phi)\ge J^h(\sigma',\tau)$ for every $\tau\in\cT$.}
\end{definition}

In view of the regularity of the wealth process and the process $J^h (\sigma',\,\cdot\,)$, the condition appearing in Definition \ref{ratioca} can be restated as follows: the inequality $V_{\sigma' \wedge t}(x_1+p^h(x_1),\phi)\ge J^h(\sigma',t)$ holds for all $t \in [0,T]$.

Assume that the game contract $\cCg$ is traded at time $0$ at the hedger's acceptable price $p^h(x_1)$. Then, from Theorem \ref{the5.1} and Propositions \ref{pro5.1} and \ref{pro5.2}, we know that there exists a triplet $(\phi',\sigma',\tau')\in\Psi(x_1+p^h(x_1),A)\times\cT\times\cT$ such that, on the one hand, $(p^h(x_1),\phi',\sigma') \in$ (SH) and, on the other hand, $(p^h(x_1),\phi',\sigma',\tau') \in$ (BE). We thus see that the classes of all hedger's rational cancellation and break-even times (see Definition \ref{breakh}) are non-empty. We first aim to characterize the class of all hedger's rational cancellation times. To this end,
we will need the following version of the comparison property for solutions to BSDEs with generator $g$.

\begin{assumption} \label{assBSDEcp}
{\rm The following {\it extended comparison property} for solutions to BSDEs holds: if for $j=1,2$
\[
\left\{ \begin{array} [c]{ll}
-dY_s^j=g_j(s,Y^j_s,Z^j_s)\,ds-Z_s^j\,dS_s+d\hHh_s^j,\medskip\\ Y^j_\tau=\zeta_j,
\end{array} \right.
\]
where $\tau\in\mathcal{T}$, $\zeta_1\ge\zeta_2,\,g_1(s,Y^2_s,Z^2_s)\geq g_2(s,Y^2_s,Z^2_s)$ for all $s\in[0,\tau]$ and the process $\hHh^1-\hHh^2$ is nondecreasing, then $Y_s^1\geq Y_s^2$ for every $s\in[0,\tau]$.}
\end{assumption}

\begin{lemma} \label{lemmh3.3}
Let Assumption \ref{assBSDEcp} be satisfied and the strict comparison property of the hedger's $g$-evaluation hold. Assume that $(Y,Z,\Lll,\Uuu)$ is the unique solution to the DRBSDE \eqref{hDRBSDE}.
If $\sigma' \in \cT$ is a hedger's rational cancellation time and $\tau'\in\cT$ is such that $\Lll_{\tau'}=0$ and $Y_{\tau'}=X^l_{\tau'}$,
then we have $\Uuu_{\sigma'\wedge \tau'}=0$ and $Y_{\sigma'\wedge\tau'}=J^h(\sigma',\tau')$.
\end{lemma}

\proof
Assume that $\sigma' \in \cT$ is a hedger's rational cancellation time for $\cCg$. Then there exists a hedger's trading strategy $\varphi\in\Psi(x_1+p^h(x_1),A)$ such that $V_{\sigma'\wedge\tau}(x_1+p^h(x_1),\varphi )\geq J^h(\sigma',\tau)$ for every counterparty's exercise time $\tau\in\cT$. The comparison property of the hedger's $g$-evaluation yields
\[
x_1+p^h(x_1)=\cEgh_{0,\sigma'\wedge\tau} \big(V_{\sigma'\wedge\tau}(x_1+p^h(x_1),\varphi)\big)\ge \cEgh_{0,\sigma'\wedge\tau} (J^h(\sigma',\tau))
\]
and thus also
\[
x_1+p^h(x_1) \ge \sup_{\tau\in\cT}\cEgh_{0,\sigma'\wedge\tau} (J^h(\sigma',\tau)).
\]
In particular, if $\tau'\in\cT$ is such that $Y_{\tau'}=X^l_{\tau'}$, then
\begin{equation}\label{eqh2g}
x_1+p^h(x_1)\ge \sup_{\tau\in\cT}\cEgh_{0,\sigma'\wedge\tau} (J^h(\sigma',\tau))\ge
\cEgh_{0,\sigma'\wedge\tau'} (J^h(\sigma',\tau'))\ge \cEgh_{0,\sigma'\wedge\tau'} (Y_{\sigma'\wedge\tau'}),
\end{equation}
where the last inequality holds because it is possible to show that $J^h(\sigma',\tau')\ge Y_{\sigma'\wedge\tau'}$. Indeed, since the quadruplet $(Y,Z,\Lll,\Uuu)$ solves the DRBSDE \eqref{hDRBSDE}, we have $Y_{\sigma'\wedge\tau'}=Y_{\tau'}=X_{\tau'}^l$ if $\tau'\le \sigma'$,
$Y_{\sigma'\wedge\tau'}=Y_{\sigma'}\leq X_{\sigma'}^u$ if $\sigma'<\tau'\leq T$ and
$Y_{\sigma'\wedge\tau'}=X_{T}^m$ if $\tau'=\sigma'=T$.
Since
\[
J^h(\sigma',\tau')=X^l_{\tau'}\I_{\{\tau'<\sigma'\}}+X^u_\sigma\I_{\{\sigma'<\tau'\}}+X^m_{\sigma}\I_{\{\tau'=\sigma'\}},
\]
$X^l< X^u$ and $X^l\leq X^m\leq X^u$, it is now easy to see that $J^h(\sigma',\tau')\ge Y_{\sigma'\wedge\tau'}$.
Recalling that $p^h(x_1)=Y_0-x_1$, we obtain from \eqref{eqh2g}
\begin{equation}\label{eqh1g}
Y_0\ge\cEgh_{0,\sigma'\wedge\tau'} (Y_{\sigma'\wedge\tau'}).
\end{equation}
By assumption about $\tau'$, we also have that $\Lll_{\tau'}=0$ holds and thus the DRBSDE \eqref{hDRBSDE} for $0\leq r\leq t\leq \tau'$ can be written as
\[
\left\{ \begin{array} [c]{ll}
-dY_r=g(r,Y_r,Z_r)\,dr-Z_r\,dS_r-dA_t-d\Uuu_r,\quad Y_t=Y_t ,\medskip\\
X^l_r \leq Y_r \leq X^u_r,\quad \int_0^t (X^u_r-Y_r)\,d\Uuu^c_r=0,\quad \Delta \Uud=\Delta (Y-A) \I_{\{ Y_{-}= X^u_{-}\}}.
\end{array} \right.
\]
Using Assumption \ref{assBSDEcp}, we obtain the inequality $\cEgh_{s,t}(Y_{t})\ge Y_s$ for $0\leq s\leq t\leq \tau'$, which means that the process $Y$ is an $\cEgh$-submartingale on $[0,\tau']$. From \eqref{eqh1g} and the postulated strict comparison property of the hedger's $g$-evaluation, we will now deduce that for every $0\leq s\leq \sigma'\wedge\tau'$
\begin{equation}\label{equationholdermore1 GCC}
\cEgh_{s,\sigma'\wedge\tau'}(Y_{\sigma'\wedge\tau'})=Y_s.
\end{equation}
Suppose, on the contrary, that equality \eqref{equationholdermore1 GCC} is not true. Then the strict comparison property of the hedger's $g$-evaluation would yield
\[
\cEgh_{0,\sigma'\wedge\tau'}(Y_{\sigma'\wedge\tau'})=\cEgh_{0,s}(\cEgh_{s,\sigma'\wedge\tau'}
(Y_{\sigma'\wedge\tau'}))>\cEgh_{0,s}(Y_s)\ge Y_0 ,
\]
which would contradict \eqref{eqh1g}. From \eqref{equationholdermore1 GCC}, we have $\cEgh_{t,\sigma'\wedge\tau'}(Y_{\sigma'\wedge\tau'})=Y_t$
and thus, for all $0\leq s\leq t\leq \sigma'\wedge\tau'$,
\[
\cEgh_{s,t}(Y_{t})=\cEgh_{s,t}(\cEgh_{t,\sigma'\wedge\tau'}(Y_{\sigma'\wedge\tau'}))
=\cEgh_{s,\sigma'\wedge\tau'}(Y_{\sigma'\wedge\tau'})=Y_s,
\]
where the last equality also comes from \eqref{equationholdermore1 GCC}. We have thus shown that $Y$ is an $\cEgh$-martingale on $[0,\sigma'\wedge\tau']$ so that $\Uuu_{\sigma'\wedge\tau'}=0$ and $\Uuu_{\sigma'\wedge t}=0$
for all $t\in[0,\tau']$.  In particular, we have
\begin{equation} \label{xeqh2g}
\cEgh_{0,\sigma'\wedge\tau'}(Y_{\sigma'\wedge\tau'})=Y_0 .
\end{equation}
By combining \eqref{eqh2g} with \eqref{xeqh2g}, we get
\begin{align*}
Y_0&=x_1+p^h(x_1)=\cEgh_{0,\sigma'\wedge\tau'} \big(V_{\sigma'\wedge\tau'}(x_1+p,\varphi)\big)
=\sup_{\tau\in\cT}\cEgh_{0,\sigma'\wedge\tau} (J^h(\sigma',\tau))
\\&=\cEgh_{0,\sigma'\wedge\tau'}(J^h(\sigma',\tau'))=\cEgh_{0,\sigma'\wedge\tau'}(Y_{\sigma'\wedge\tau'}).
\end{align*}
Since we have shown that $J^h(\sigma',\tau')\ge Y_{\sigma'\wedge\tau'}$, using the strict comparison property of
the hedger's $g$-evaluation, we obtain the desired equality $J^h(\sigma',\tau')=Y_{\sigma'\wedge\tau'}$.
\finproof

We will study two possible exercise times for the counterparty
\[
 \tau^h:=\inf\{t \in [0,T]\,|\,Y_t=X^l_t\}, \quad \bar\tau^h:=\inf\{t \in [0,T]\,|\,\Lll_t>0\},
\]
which are of a special interest for the hedger's valuation and rational cancellation problems, as emphasized by
the superscript $h$.

\begin{proposition} \label{proh3.3}
Let Assumptions \ref{assRBSDEz}--\ref{assBSDEcp} be satisfied and the strict comparison property of the hedger's $g$-evaluation hold. Then the following assertions are true:\\
(i) if a stopping time $\sigma'\in \cT$ is such that (a) $\Uuu_{\sigma'}=0$ and (b) $Y_{\sigma'}=X^u_{\sigma'}$, then $\sigma'$  is a hedger's rational cancellation time,  \\
(ii) if $\sigma'$ is a hedger's rational cancellation time, then $\Uuu_{\sigma'\wedge \tau^h}=\Uuu_{\sigma'\wedge \bar\tau^h}=0$, $Y_{\sigma'\wedge\tau^h}=J^h(\sigma',\tau^h)$ and $Y_{\sigma'\wedge \bar\tau^h}=J^h(\sigma',\bar\tau^h)$.
\end{proposition}

\proof (i) If condition (a) is satisfied then, arguing as in the proof of Proposition \ref{pro5.1}, we obtain the existence of a hedger's trading strategy $\phi \in \Psi (x_1+p^h(x_1),A)$ such that the wealth process $V=V(x_1+p^h(x_1),\phi,A)$ satisfies $V_t \ge Y_t \ge X^l_t$ for all $t \in [0,\sigma']$. Moreover, since (b) holds, we get $V_{\sigma'}\ge  Y_{\sigma'}=X^u_{\sigma'} \ge X^m_{\sigma'}$ and, for all $t\in [0,T]$,
\[
V_{\sigma' \wedge t}\ge X^l_t\I_{\{t<\sigma'\}}+X^u_{\sigma'}\I_{\{\sigma'<t\}}+X^m_{\sigma'}\I_{\{\sigma'=t\}}=J^h(\sigma',t).
\]
It is thus clear that $\sigma'$ is a hedger's rational cancellation time for $\cCg$.

\noindent (ii) In view of Lemma \ref{lemmh3.3}, it suffices to show that $\Lll_{\tau^h}=\Lll_{\bar\tau^h}=0,\, Y_{\tau^h}=X^l_{\tau^h}$ and $Y_{\bar\tau^h}=X_{\bar\tau^h}^l$. The equality $Y_{\tau^h}=X^l_{\tau^h}$ follows from the definition of $\tau^h$ and the right-continuity of processes $X^l$ and $Y$. Hence, in view of \eqref{hDRBSDE}, from the continuity of $L$ we obtain $\Lll_{\tau^h}=0$. We will now show that $\Lll_{\bar\tau^h}=0$ and $Y_{\bar\tau^h}=X_{\bar\tau^h}^l$. For an arbitrary $\varepsilon>0$, there exists a $\delta\in[0,\varepsilon]$ such that $\Lll_{\bar\tau^h+\delta}>0$ and thus, since $\int_0^T (Y_t-X_t^l)\, d\Lll_t=0$, there exists a $\delta_1\in[0,\delta]$ such that $Y_{\bar\tau^h+\delta_1}=X^l_{\bar\tau^h+\delta_1}$. From the right-continuity of $Y$ and $X^l$ and the fact that $\varepsilon$ and $0\leq\delta_1\leq\delta\leq\varepsilon$ are arbitrary, we deduce that $Y_{\bar\tau^h}=X_{\bar\tau^h}^l$. Moreover, since $\Lll_t=0$ for all $t\in[0,\bar\tau^h)$, in view of the continuity of $L$, we also have that $\Lll_{\bar\tau^h}=0$.
\finproof

\begin{corollary} \label{corr3.3}
Let Assumptions \ref{assRBSDEz}--\ref{assBSDEcp} be satisfied and the strict comparison property of the hedger's $g$-evaluation hold. Then the stopping times $\sigma^h$ and $\bar{\sigma}^h$, which are given by
\begin{equation*} 
 \sigma^h:=\inf\{t\in [0,T]\,|\,Y_t=X^u_t\}, \quad \bar{\sigma}^h:=\inf\{t\in [0,T]\,|\, \Uuu_t >0 \},
\end{equation*}
are hedger's rational cancellation times.
\end{corollary}

\proof We first consider the stopping time $\sigma^h$.  From the right-continuity of $Y$ and $X^u$, we obtain the equality $Y_{\sigma^h}=X^u_{\sigma^h}$. Moreover, from the definition of $\sigma^h$, we have $Y_t<X_t^u$ for $t\in[0,\sigma^h)$ and thus $\Uuu_t=0$ for all $t\in[0,\sigma^h]$ observing that $U$ is continuous. Using part (i) in Proposition \ref{proh3.3}, we conclude that $\sigma^h$ is a hedger's rational cancellation time. Let us now focus on the stopping time $\bar \sigma^h$. Then, similarly to the proof of Proposition \ref{proh3.3}, using the right-continuity of $Y$ and $X^u$, we can show that $Y_{\bar \sigma^h}=X_{\bar \sigma^h}^u$ and $\Uuu_t=0$ for all $t\in[0,\bar \sigma^h]$. Thus, using again part (i) in Proposition \ref{proh3.3}, we conclude that $\bar \sigma^h$ is also one of hedger's rational cancellation times.
\finproof

In the case of an American option and a nonlinear market, it is easy to characterize the earliest and latest holder's rational exercise times for the holder of the option (see, for instance, Section 3.6 in Kim et al. \cite{KNR2018}). In contrast, since a game contract can be stopped by either of the two parties at any instance, it is much harder to provide a full characterization of the earliest and latest rational cancellation times for the hedger and, by the same token, the earliest and latest rational exercise times for the counterparty. Nevertheless, it is possible to show that the hedger's rational cancellation times $\sigma^h$ and $\bar{\sigma}^h$ enjoy the property of being, at least in some special circumstances, the earliest and the latest among all hedger's rational cancellation times.

\begin{corollary} \label{corr3.4}
Let Assumptions \ref{assRBSDEz}--\ref{assBSDEcp} be satisfied and the strict comparison property of the hedger's $g$-evaluation hold. Then the following assertions are true:\\
(i)  if $\sigma'$ is a hedger's rational cancellation time such that $\sigma' \leq \sigma^h$ on the event $E^h:=\{ \sigma^h \leq \bar{\tau}^h\}$, then $\sigma' = \sigma^h$ on $E^h$, \\
(ii) if $\sigma'$ is a hedger's rational cancellation time such that $\sigma' \geq \bar{\sigma}^h$ on the event  $\bar{E}^h:=\{ \bar{\sigma}^h < \bar{\tau}^h\}$, then $\sigma' = \bar{\sigma}^h$ on  $\bar{E}^h$.
\end{corollary}

\proof
(i) We argue by contradiction. Let $\sigma'$ be a stopping time such that $\sigma' \leq \sigma^h$ on the event $E^h$ and
$\P ( \{\sigma' < \sigma^h\} \cap E^h )>0$. In view of the assumption that $\bar\tau^h \geq \sigma^h$ on $E^h$, we have $\sigma'\wedge\bar\tau^h=\sigma'$ on $E^h$. Due to the definition of the stopping time $\sigma^h$, we have that $Y_{\sigma'}<X^u_{\sigma'}$ on  $\{\sigma' < \sigma^h\}$. Thus $Y_{\sigma'\wedge\bar\tau^h}=Y_{\sigma'}<X^u_{\sigma'}=X^u_{\sigma'\wedge\bar\tau^h}=J^h(\sigma',\bar\tau^h)$ on
$\{\sigma' < \sigma^h\} \cap E^h$ where the last equality follows from \eqref{eq.5.0}. In view of part (ii) in Proposition \ref{proh3.3}, we conclude that $\sigma'$ cannot be a hedger's rational cancellation time.

\noindent (ii) Once again we argue by contradiction. Let $\sigma'$ be a stopping time such that $\sigma' \geq \bar{\sigma}^h$ on the event $\bar{E}^h$ and $\P (\{\sigma' > \bar{\sigma}^h\} \cap \bar{E}^h)>0$. Since manifestly $\bar\tau^h > \bar\sigma^h$ on $\bar{E}^h$, we have that $\P (\sigma'\wedge \bar\tau^h> \bar\sigma^h)>0$. Due to the definition of $\bar\sigma^h$, this means that $\P (U_{\sigma'\wedge \bar\tau^h}>0)>0$ and thus, in view of part (ii) in Proposition \ref{proh3.3}, it is clear that $\sigma'$ cannot be a hedger's rational cancellation time.
\finproof

From Corollary \ref{corr3.4} it follows, in particular, that if the inequality $\bar\tau^h \geq \sigma^h$ holds, then $\sigma^h$ is  the earliest among all hedger's rational cancellation times, that is, if $\sigma'$ is any hedger's rational cancellation time such that $\sigma' \leq \sigma^h$, then $\sigma' = \sigma^h$. Similarly, if $\bar\tau^h > \bar\sigma^h$, then $\bar{\sigma}^h$ is the latest among all hedger's rational cancellation times, that is, if $\sigma'$ is any hedger's  rational cancellation time such that $\sigma' \geq \bar{\sigma}^h$, then the equality $\sigma' = \bar{\sigma}^h$ holds.

\subsection{Hedger's Break-Even Times} \label{sec3.4c}

In the next step we will examine the set of {\it hedger's break-even times}, which are introduced in Definition~\ref{breakh}.
Obviously, a hedger's break-even time is in fact a particular exercise time selected by the counterparty, but its name is aimed to emphasize that the choice of that time by the counterparty affects in a very special way the hedger's financial outcome.
In other words, it is a particular exercise time related to the hedger's unilateral valuation problem, but it is usually meaningless
for the counterparty's unilateral valuation and exercising problems in a general nonlinear framework.

\begin{definition}  \label{breakh}
{\rm A stopping time $\tau\in\cT$ is called a {\it hedger's break-even time} for the triplet $(p,\phi,\sigma)\in\rr\times\Psi(x_1 + p,A)\times\cT$ if condition (BE) is satisfied by the quadruplet $(p,\phi,\sigma,\tau)$.}
\end{definition}

\begin{remark} {\rm Using the symmetry of the problem, we will also analyze in Section \ref{sec3.5b} the counterparty's rational exercise time, which is chosen by the counterparty, and the counterparty's break-even time, which is selected by the hedger, but is relevant for the financial outcome of the counterparty.}
\end{remark}

In the next result, we provide several alternative characterizations of all hedger's break-even times associated with the hedger's replicating strategy $(p^h(x_1),\phi',\sigma')=(Y_0-x_1,Z,\sigma^h)$. Note that the stopping time $\sigma' = \sigma^h$ is a hedger's rational cancellation time and $\phi'=Z$ where the quadruplet $(Y,Z, \Lll, \Uuu)$ is a unique solution to the DRBSDE (\ref{hDRBSDE}).
Recall that the stopping time $\tau^h$ is given by $\tau^h :=\inf\{t\in [0,T]\,|\,Y_t=X^l_t\}$.

\begin{proposition} \label{proh3.4}
Let Assumption \ref{assBSDEcp} be satisfied and $(Y,Z, \Lll, \Uuu)$ be the unique solution to the hedger's DRBSDE (\ref{hDRBSDE}) where the process $-X^u+A$ is assumed to be left-upper-semicontinuous.
For $(p^h(x_1),\phi',\sigma')=(Y_0-x_1,Z,\sigma^h)$, the following assertions are equivalent: \\
(i)  a stopping time  $\tau'\in\cT$ is a hedger's break-even time for the triplet $(p^h(x_1),\phi',\sigma')\in\rr\times\Psi(x_1 + p^h(x_1),A)\times\cT$,\\
(ii) the quadruplet $(p^h(x_1),\phi',\sigma',\tau')\in\rr\times\Psi(x_1 + p^h(x_1),A)\times\cT\times\cT$ satisfies condition (NA),\\
(iii) the equality $V_{\sigma'\wedge\tau'}(x_1+p^h(x_1),\phi')=J^h(\sigma',\tau')$ holds,\\
(iv) the equalities $Y_{\sigma'\wedge\tau'}=J^h(\sigma'\wedge\tau')$ and $\Lll_{\tau'\wedge\sigma'}=\Uuu_{\sigma'}=0$ hold and thus the process $Y$ is an $\cEgh$-martingale on $[0,\sigma'\wedge\tau']$,\\
(v) a stopping time $\tau' \in \cT$ is a solution to the following nonlinear optimal stopping problem: find $\tau' \in \cT$ such that
\[
\cEgh_{0,\sigma'\wedge\tau'}\big(J^h(\sigma',\tau')\big)=\sup_{\tau\in\cT}\cEgh_{0,\sigma'\wedge\tau}(J^h(\sigma',\tau)).
\]
Moreover, if the process $X^l-A$ is left-upper-semicontinuous, then $\tau^h$ is a hedger's break-even time for the triplet $(p^h(x_1),\phi',\sigma')$. If, in addition, the inequality $\sigma'\ge\tau^h$ holds, then $\tau^h$ is the earliest hedger's break-even time for the triplet $(p^h(x_1),\phi',\sigma')$.
\end{proposition}

\proof
From results established in Section \ref{sec2.3}, we note that since $(p^h(x_1),\phi',\sigma')$ is a hedger's replicating strategy (see Proposition  \ref{pro5.2}), it is also his superhedging strategy, and thus it is clear that assertions (i), (ii) and (iii) are indeed equivalent.

\noindent  (iii)$\, \Rightarrow\, $(iv). From the proof of Proposition \ref{pro5.1}, we know that
$V_t(x_1+p^h(x_1),\phi' )\geq Y_t \geq X_t^l $ for all $t\in[0,\sigma']$ and thus, in particular,
$V_{\sigma' \wedge \tau'}(x_1+p^h(x_1),\phi') \ge Y_{\sigma' \wedge \tau'}$. Moreover,
since  $(Y,Z, \Lll, \Uuu)$ solves the DRBSDE (\ref{hDRBSDE}), we have $X^l\leq Y\leq X^u$ and $Y_{\sigma'}=X_{\sigma'}^u\ge X_{\sigma'}^m $. Therefore,
\begin{equation}\label{inequality GCC}
V_{\sigma'\wedge\tau'}(x_1+p^h(x_1),\phi')\ge Y_{\sigma'\wedge\tau'}\ge X^l_{\tau'}\I_{\{\tau'<\sigma'\}}+X^u_{\sigma'}\I_{\{\sigma'<\tau'\}}
+X^m_{\sigma'}\I_{\{\sigma'=\tau'\}}=J^h(\sigma',\tau').
\end{equation}
From (iii), $V_{\sigma'\wedge\tau'}(x_1+p^h(x_1),\phi')=Y_{\sigma'\wedge\tau'}=J^h(\sigma',\tau')$, which is $V_{\sigma'\wedge\tau'}(Y_0,\phi')
=Y_{\sigma'\wedge\tau'}=J^h(\sigma',\tau')$ by recalling that $p^h(x_1)=Y_0-x_1$. Moreover, since the process $V=V(Y_0,\phi' )$ satisfies the SDE (\ref{eq.fSDE2}), it is easy to see that the process $V(Y_0,\phi')$ is an $\cEgh$-martingale. We thus have
\begin{equation}\label{equation1g}
\cEgh_{0,\sigma'\wedge\tau'}(Y_{\sigma'\wedge\tau'})=\cEgh_{0,\sigma'\wedge\tau'}\big(V_{\sigma'\wedge\tau'}(Y_0,\phi' )\big)=Y_0.
\end{equation}
The process $Y$ solves the DRBSDE \eqref{hDRBSDE}, which can be written on $0\leq r\leq t\leq\sigma'$ in the following way
(recall that $\Uuu_{\sigma'}=0$)
\[
\left\{ \begin{array} [c]{ll}
dY_r=-g(r,Y_r,Z_r)\,dr+Z_r\,dS_r+dA_t-d\Lll_r ,\medskip\\
Y_t=Y_t,\quad X^l_r \leq Y_r \leq X^u_r,\quad \int_0^t (Y_r-X^l_r)\,d\Lll^c_r=0,\quad \Delta \Lld=-\Delta (Y-A) \I_{\{ Y_{-}=X^l_{-}\}}.
\end{array} \right.
\]
Using Assumption \ref{assBSDEcp}, we obtain $\cEgh_{s\wedge\sigma',t\wedge\sigma'}(Y_{t\wedge\sigma'})\leq Y_{s\wedge\sigma'}$ for all $0\leq s\leq t\leq T$, so that $Y$ is an $\cEgh$-supermartingale on the stochastic interval $[0,\sigma']$. We now claim that, for all $0\leq s\leq \tau'$,
\begin{equation}\label{equationmore1 GCC}
\cEgh_{s\wedge\sigma',\tau'\wedge\sigma'}(Y_{\tau'\wedge\sigma'})=Y_{s\wedge\sigma'}.
\end{equation}
Suppose, on the contrary, that the equality \eqref{equationmore1 GCC} fails to hold. In that case, the strict comparison property of the hedger's $g$-evaluation would yield
\[
\cEgh_{0,\tau'\wedge\sigma'}(Y_{\tau'\wedge\sigma'})=
\cEgh_{0, s\wedge\sigma'}(\cEgh_{s\wedge\sigma',\tau'\wedge\sigma'}(Y_{\tau'\wedge\sigma'}))
<\cEgh_{0, s\wedge\sigma'}(Y_{s\wedge\sigma'})\leq Y_0 ,
\]
which manifestly contradicts \eqref{equation1g}. Next, we claim that $\cEgh_{s\wedge\sigma', t\wedge\sigma'}(Y_{t\wedge\sigma'})=Y_{s\wedge\sigma'}$ for $0\leq s\leq t\leq T$, which means that $Y$ is an $\cEgh$-martingale on $[0,\tau'\wedge\sigma' ]$ and thus $\Lll_{\tau'\wedge\sigma' }=0$. To establish this property,
we note that \eqref{equationmore1 GCC} gives $\cEgh_{t\wedge\sigma',\tau'\wedge\sigma'}(Y_{\tau'\wedge\sigma'})=Y_{t\wedge\sigma'}$
and thus, for all $0\leq s\leq t\leq T$,
\[
\cEgh_{s\wedge\sigma', t\wedge\sigma'}(Y_{t\wedge\sigma'})
=\cEgh_{s\wedge\sigma', t\wedge\sigma'}(\cEgh_{t\wedge\sigma',\tau'\wedge\sigma'}(Y_{\tau'\wedge\sigma'}))
=\cEgh_{s\wedge\sigma',\tau'\wedge\sigma'}(Y_{\tau'\wedge\sigma'})=Y_{s\wedge\sigma'},
\]
where the last equality comes from  \eqref{equationmore1 GCC}.

\noindent  (iv)$\, \Rightarrow\, $(iii).  We observe that $Y_{\sigma' \wedge \tau'}=J^h(\sigma' \wedge \tau')$, $\Lll_{\tau'\wedge\sigma'}=\Uuu_{\sigma'}=0$ and $(Y,Z, \Lll, \Uuu)$ solves the DRBSDE (\ref{hDRBSDE}), which thus reduces to the following BSDE on the stochastic interval $[0,\sigma'\wedge\tau']$
\[
\left\{ \begin{array} [c]{ll}
-dY_r=g(r,Y_r,Z_r)\,dr-Z_r\,dS_r-dA_t,\medskip\\
Y_{\sigma'\wedge\tau' }=J^h(\sigma' \wedge \tau').
\end{array} \right.
\]
Given the process $Z$, the above BSDE can be formally rewritten as the (forward) SDE, for all $r\in[0,\tau']$,
\[
\left\{ \begin{array} [c]{ll}
dY_r=-g(r,Y_r,Z_r)\,dr+Z_r\,dS_r+dA_t,\medskip\\
Y_{0}=Y_{0}.
\end{array} \right.
\]
Recall that $V(x_1+ p^h(x_1),\phi' )=V(Y_0,Z)$ satisfies the following SDE,  for all $r\in[0,T]$,
\[
\left\{ \begin{array} [c]{ll}
dV_r=-g(r,V_r,Z_r)\,dr+Z_r\,dS_r+dA_t,\medskip\\
V_{0}=Y_{0}.
\end{array} \right.
\]
From the uniqueness of a solution to the SDE, we deduce that $V(Y_0,Z)$ and $Y$ coincide on $[0,\sigma'\wedge\tau']$.
In particular, $V_{\sigma' \wedge \tau'}(x_1+ p^h(x_1),\phi' )=Y_{\sigma'\wedge\tau'}=J^h(\sigma' \wedge \tau')$.

\noindent  (iv)$\, \Leftrightarrow\, $(v).  First, from (iv) we know that
$\cEgh_{0,\sigma'\wedge\tau' }(J^h(\sigma' \wedge \tau'))=\cEgh_{0,\sigma'\wedge\tau' }(Y_{\sigma'\wedge\tau' })=Y_0$.
Observe that Assumption \ref{assDGw} yields
$Y_0=\sup_{\tau\in\cT } \cEgh_{0,\sigma'\wedge\tau }(J^h(\sigma' \wedge \tau))$ and thus
\[
\cEgh_{0,\sigma'\wedge\tau' }(J^h(\sigma' \wedge \tau'))=\sup_{\tau\in\cT } \cEgh_{0,\sigma'\wedge\tau }(J^h(\sigma' \wedge \tau)).
\]
Conversely, if (v) holds, then Assumption \ref{assDGw} gives
\[
Y_0=\cEgh_{0,\sigma'\wedge\tau' }(J^h(\sigma' \wedge \tau')) \leq\cEgh_{0,\sigma' \wedge\tau' }(Y_{\sigma'\wedge\tau' })
\]
where the last inequality is a consequence of (\ref{inequality GCC}). Thus, similarly as for the implication (iii)$\Rightarrow$(iv), one can show (iv) holds. This completes the proof of the equivalence (iv)$\, \Leftrightarrow\, $(v).

Let us show that the stopping time $\tau^h$ is a hedger's break-even time. From the right-continuity of $Y$ and $X$, we deduce that $Y_{\tau^h}=X_{\tau^h}^l$. Moreover, from the definition of $\tau^h$, we have $Y_t>X_t^l$ for $t\in[0,\tau^h)$ and thus $\Lll_t=0$ for all  $t\in[0,\tau^h]$. Therefore, we have $Y_{\tau^h}=X_{\tau^h}^l$ and $\Lll_{\tau^h}=0$, that is, (iv) holds with $\tau'=\tau^h$. Hence, from the above equivalences (in particular, (iv) $\Leftrightarrow$ (i)), we infer that $\tau^h$ is a hedger's break-even time for the triplet $(p^h(x_1),\phi',\sigma')\in\rr\times\Psi(x_1 + p^h(x_1),A)\times\cT$. One can also provide another simple argument: from Assumption \ref{assDGw}, we have that $\tau^h$ is a solution to the following nonlinear optimal stopping problem:
find $\tau' \in \cT$ such that
\[
\cEgh_{0,\sigma'\wedge\tau'}\big(J^h(\sigma',\tau^h)\big)=\sup_{\tau\in\cT}\cEgh_{0,\sigma'\wedge\tau}(J^h(\sigma',\tau))
\]
and, from the equivalence (v) $\Leftrightarrow$ (i), we infer that $\tau^h$ is a hedger's break-even time for the triplet $(p^h(x_1),\phi',\sigma')$.

To complete the proof of the proposition, it remains to show that if, in addition, the inequality $\sigma'\ge\tau^h$ holds, then $\tau^h$ is the earliest hedger's break-even time for the triplet $(p^h(x_1),\phi',\sigma')$.
We argue by contradiction. Let $\tau'$ be any hedger's break-even time for the triplet $(p^h(x_1),\phi',\sigma')$ such that $\tau'\leq \tau^h$  and $\P ( \{\tau' < \tau^h\} )>0$. Then, from (iv) and  $\sigma'\ge\tau^h\ge\tau'$, it holds that $Y_{\tau'}=X_{\tau'}^l$ on $\{\tau' < \tau^h\}$, which clearly contradicts the definition of $\tau^h$.
\finproof

\subsection{Counterparty's Acceptable Price via DRBSDE} \label{sec3.5}

In Sections \ref{sec3.5} and \ref{sec3.5b}, it is postulated without further mentioning that Assumptions \ref{assnBSDE}--\ref{assDGw} are satisfied by the counterparty's wealth processes associated with the process $-A$ and the counterparty's DRBSDE \eqref{cDRBSDE}.
Observe that the the counterparty's DRBSDE is specified in the manner analogous to the hedger's equation \eqref{hDRBSDE}
and the counterparty's $g$-evaluation $\cEgc$ is defined in the same way as the hedger's $g$-evaluation $\cEgh$ but with the process $A$ replaced by the process $-A$ in the BSDE \eqref{nhBSDE}.

In particular, the lower and upper obstacles in the counterparty's DRBSDE are  c\`adl\`ag processes $x^l$ and $x^u$ given by $x^l_t:=X^h_t+\Vben_t(x_2)<x^u_t:=X^c_t+\Vben_t(x_2)$ for all $t \in [0,T]$ and the terminal value $x^m_T$ is such that $x^l_T \leq x^m_T:=\Xxb_T+\Vben_T(x_2) \leq x^u_T$. An application of Assumption \ref{assRBSDEh} to the counterparty's  DRBSDE ensures the existence of a unique solution $(y,z,\cLll,\cUuu)$ to the DRBSDE \eqref{cDRBSDE}  with parameters $(g,x^l,x^u,x^m_T)$
\begin{equation} \label{cDRBSDE}
\left\{ \begin{array} [c]{ll}
-dy_t=g(t,y_t,z_t)\,dt-z_t\,dS_t+dA_t+ d\cLll_t-d\cUuu_t,\quad y_T=x^m_T,\medskip\\
x^l_t \leq y_t \leq x^u_t,\quad \int_0^T (y_t-x^l_t)\,d\cLll^c_t=\int_0^T (x^u_t-y_t)\,d\cUuu^c_t=0,\medskip\\
\Delta \cLld=-\Delta (y+A) \I_{\{ y_{-}= \zeta^l_{-}\}},\quad \Delta \cUud=\Delta (y+A) \I_{\{ y_{-}= \zeta^u_{-}\}},
\end{array} \right.
\end{equation}
where, in particular, $\cLll$ and $\cUuu$ are $\gg$-predictable, c\`adl\`ag, nondecreasing processes such that $\cLll_0=\cUuu_0=0$. Furthermore, $\cLll=\cLlc+\cLld$ and $\cUuu=\cUuc+\cUud$ give their unique decompositions into continuous and jump components.

Recall also that the counterparty's relative reward $J^c(\sigma ,\tau)= \wt{J}(x_1,x^l,x^u,x^m,\sigma,\tau)$ is given by equation \eqref{eq4.5h}, that is,
\[
J^c(\sigma,\tau)=I(\sigma,\tau)+\Vben_{\sigma \wedge \tau }(x_2)=
x^l_\sigma\I_{\{\sigma<\tau\}}+x^u_{\tau}\I_{\{\tau<\sigma\}}+x^m_{\sigma}\I_{\{\tau=\sigma\}}.
\]

In Sections \ref{sec3.4} and \ref{sec3.4b}, we have established several results for the hedger.
Since the two unilateral valuation and hedging problems are in some sense symmetric, it is clear that the analogous results for the counterparty should be valid as well and thus it suffices to give their statements without proofs.
We first state the counterparty's version of Proposition \ref{pro5.1}, which furnishes the link between the counterparty's superhedging costs and the unique solution $(y,z,\cLll,\cUuu)$ to the counterparty's DRBSDE~\eqref{cDRBSDE}.

\begin{proposition}   \label{pro5.4c}
Let the process $-x^u-A$ be left-upper-semicontinuous. If the comparison property of the counterparty's $g$-evaluation holds, then the upper bound for the counterparty's superhedging costs satisfies
\[
\psup^{s,c}(x_2)=x_2-y_0=x_2 - \inf_{\tau\in\cT}\sup_{\sigma\in\cT}\cEgc_{0,\sigma \wedge \tau}(J^c (\sigma,\tau)).
\]
\end{proposition}

In the next result, which corresponds to Proposition \ref{pro5.2} and gives a counterparty's replicating strategy for
$\cCg$, we will need the following analogue of Assumption \ref{assRBSDEz}.

\begin{assumption} \label{assRBSDEcz}
{\rm  The processes $x^l+A$ and $-x^u-A$ are left-upper-semicontinuous so that the processes $\cLll$ and $\cUuu$
 in the solution $(y,z, \cLll, \cUuu)$ to the counterparty's DRBSDE \eqref{cDRBSDE} are continuous.}
\end{assumption}

We define the hedger's cancellation time $\sigma^c$ and the counterparty's exercise time $\tau^c$ by setting
\[
\tau^c:=\inf\{t \in [0,T]\,|\,y_t=x^u_t\}, \quad \sigma^c:=\inf\{t\in [0,T]\,|\,y_t=x^l_t\}.
\]

\begin{proposition}   \label{pro5.5}
Let Assumption \ref{assRBSDEcz} be satisfied. If the comparison property of the counterparty's $g$-evaluation holds, then the following assertions are valid: \\
(i) $(x_2-y_0,z,\tau^c)$ is a counterparty's replicating strategy for $\cCg$,\\
(ii) the counterparty's maximum superhedging and replication costs satisfy
\begin{equation*}
\wh{p}^{r,c}(x_2)=\wh{p}^{s,c}(x_2) = x_2-y_0 = x_2 -\cEgc_{0,\sigma^c \wedge \tau^c }(J^c(\sigma^c,\tau^c )).
\end{equation*}
\end{proposition}

Let $\overline{\cV}^c_0(x^l,x^u,x^m)$ stand for the upper value 
for the counterparty's nonlinear Dynkin game with the payoff $J^c (\sigma,\tau)$, that is,
\begin{align*}
&\overline{\cV}^c_0(x^l,x^u,x^m):=\inf_{\tau \in\cT}\sup_{\sigma\in\cT}\cEgc_{0,\sigma\wedge\tau}(J^c (\sigma,\tau)).
\end{align*}
The following result, which is a straightforward consequence of Theorem \ref{the5.1}, shows that the counterparty's maximum replication cost coincides with his minimum fair price and furnishes alternative representations for the counterparty's acceptable price $p^c(x_2)$.

\begin{theorem}  \label{the5.2}
Let Assumption \ref{assRBSDEcz} be satisfied. If the strict comparison property of the counterparty's $g$-evaluation holds, then the unique counterparty's acceptable price $p^c(x_2)$ satisfies
\[
p^c(x_2) =\wh{p}^{f,r,c}(x_2)=\wh{p}^{r,c}(x_2)=\pmin^{f,c}(x_2)=x_2-y_0=x_2- \overline{\cV}^c_0(x^l,x^u,x^m).
\]
\end{theorem}

\subsection{Counterparty's Rational Exercise and Break-Even Times} \label{sec3.5b}

Let us state the counterparty's versions of Definition \ref{ratioca} and Proposition \ref{proh3.3}.

\begin{definition} \label{ratioex}
{\rm A stopping time $\tau'\in\cT$ is a {\it counterparty's rational exercise time} for $\cCg$ if the contract is traded at the counterparty's acceptable price $p^c(x_2)$ at time $0$ and there exists a trading strategy $\psi\in\Psi(x_2-p^c(x_2),-A)$ such that $V_{\tau'}(x_2-p^c(x_2),\psi)\ge J^c(\sigma,\tau')$ for every $\sigma\in\cT$.}
\end{definition}

The following hedger's cancellation times
\[
\sigma^c:=\inf\{t \in [0,T]\,|\,y_t=x^l_t\},\quad \bar\sigma^c:=\inf\{t \in [0,T]\,|\,\cLll_t>0\},
\]
play an important role in the counterparty's valuation problem and thus they are designated by the superscript $c$.

\begin{proposition} \label{proh3.3c}
Let Assumption \ref{assBSDEcp}--\ref{assRBSDEcz} be satisfied. If the strict comparison property of the counterparty's $g$-evaluation holds, then the following assertions are true:\\
(i) if a stopping time $\tau'\in \cT$ is such that (a) $\cUuu_{\tau'}=0$ and (b) $y_{\tau'}=x^u_{\tau'}$, then $\tau'$
is a counterparty's rational exercise time,  \\
(ii) if $\tau'\in \cT$ is a counterparty's rational exercise time, then $u_{\sigma^c\wedge \tau'}=u_{\bar\sigma^c\wedge \tau'}=0$, $Y_{\sigma^c\wedge\tau'}=J^c(\sigma^c,\tau')$ and $Y_{\bar\sigma^c\wedge\tau'}=J^c(\bar\sigma^c,\tau')$.
\end{proposition}

The following results correspond to Corollaries \ref{corr3.3} and \ref{corr3.4}, respectively.

\begin{corollary} \label{corr3.3c}
 Let Assumptions \ref{assBSDEcp}--\ref{assRBSDEcz} be satisfied. If the strict comparison property of the counterparty's $g$-evaluation holds, then the stopping times $\tau^c$ and $\bar{\tau}^c$ given by
\begin{equation*}
 \tau^c:=\inf\{t\in [0,T]\,|\,y_t=x^u_t\}, \quad \bar{\tau}^c:=\inf\{t\in [0,T]\,|\, \cUuu_t >0 \},
\end{equation*}
are counterparty's rational exercise times.
\end{corollary}

\begin{corollary} \label{corr3.4c}
 Let Assumptions \ref{assBSDEcp}--\ref{assRBSDEcz} be satisfied and the strict comparison property of the counterparty's $g$-evaluation hold. Then the following assertions are true: \\
(i)  if $\tau'$ is a counterparty's rational exercise time such that $\tau' \leq \tau^c$ on the event $E^c:=\{ \tau^c \leq \bar\sigma^c\}$, then $\tau' = \tau^c$ on $E^c$, \\
(ii) if $\tau'$ is a counterparty's rational exercise time such that $\tau' \geq \bar{\tau}^c$ on the event  $\bar{E}^c:=\{ \bar{\tau}^c < \bar\sigma^c \}$, then $\tau' = \bar{\tau}^c$ on $\bar{E}^c$.
\end{corollary}

In particular,  if $\bar\sigma^c \geq \tau^c$, then $\tau^c$ is the earliest among all counterparty's rational exercise times, that is, if $\tau'$ is any counterparty's rational exercise time such that $\tau' \leq \tau^c$, then $\tau' = \tau^c$.
Moreover, if $\bar\sigma^c > \bar\tau^c$, then $\bar{\tau}^c$ is the latest among all counterparty's rational exercise times, that is, if $\tau'$ is any counterparty's rational exercise time such that $\tau' \geq \bar{\tau}^c$, then $\tau' = \bar{\tau}^c$.

As expected, the definition of the counterparty's break-even time mimics the one for the hedger. Observe that a counterparty's
break-even time is associated with a solution to the counterparty's unilateral valuation problem (but, of course, not with
the hedger's valuation problem, in general, unless we deal with a linear market model).

\begin{definition} \label{breakc}
{\rm If condition (BE$'$) is satisfied by a quadruplet $(p,\psi,\sigma,\tau)$, then $\sigma\in\cT$ is called a {\it counterparty's break-even time} for the triplet $(p,\psi,\tau)\in\rr\times\Psi(x_2-p,-A)\times \cT$.}
\end{definition}

We conclude this work by considering the counterparty's replicating strategy $(p^c(x_2),\psi',\tau')=(x_2-y_0,z,\tau^c)$, where the quadruplet $(y,z,\cLll,\cUuu)$ is a solution to the DRBSDE (\ref{cDRBSDE}) and the stopping time $\tau' = \tau^c$ is a counterparty's rational exercise time. The proof of Proposition \ref{proh3.4c}, which provides several alternative characterizations of counterparty's break-even times associated with the triplet $(x_2-y_0,z,\tau^c)$, is exactly the same as the proof of Proposition \ref{proh3.4} and thus it is omitted.

\begin{proposition} \label{proh3.4c}
Let $(y,z,\cLll,\cUuu)$ be the unique solution to the counterparty's DRBSDE (\ref{cDRBSDE}) where the process $-x^u-A$ is assumed to be left-upper-semicontinuous. For $(p^c(x_2),\psi',\tau')=(x_2-y_0,z,\tau^c)$, the following assertions are equivalent: \\
(i) a stopping time $\sigma'\in\cT$ is a counterparty's break-even time for the triplet $(p^c(x_2),\psi',\tau')\in\rr\times\Psi(x_2-p^c(x_2),-A)\times\cT$, that is, $V_{\sigma'\wedge\tau'}(x_2-p^c(x_2),\phi' )=J^c(\sigma',\tau')$,\\
(ii) the quadruplet $(p^c(x_2),\psi',\sigma',\tau')\in\rr\times\Psi(x_2-p^c(x_2),-A)\times\cT\times\cT$ fulfills condition (NA$'$),\\
(iii) the equality $y_{\sigma'\wedge\tau'}=J^c(\sigma'\wedge\tau')$ holds,\\
(iv) the equalities $y_{\sigma'\wedge \tau'}=J^c(\sigma'\wedge\tau')$ and $\cLll_{\sigma'\wedge\tau'}=\cUuu_{\tau'}=0$ hold and thus the process $y$ is an $\cEgc$-martingale on $[0,\sigma'\wedge\tau']$,\\
(v) a stopping time $\sigma'\in\cT$ is a solution to the following nonlinear optimal stopping problem: find $\sigma' \in \cT$ such that
\[
\cEgc_{0,\sigma'\wedge\tau'}\big(J^c(\sigma',\tau')\big)=\sup_{\sigma\in\cT}\cEgc_{0,\sigma\wedge\tau'}\big(J^c(\sigma,\tau')\big).
\]
Furthermore, if the process $x^l+A$ is left-upper-semicontinuous, then $\sigma^c$ is a counterparty's break-even time for the triplet $(p^c(x_2),\psi',\tau')$. If, in addition, the inequality $\tau'\ge\sigma^c$ holds, then $\sigma^c$ is the earliest counterparty's break-even time for the triplet $(p^c(x_2),\psi',\tau')$.
\end{proposition}

\section*{Acknowledgments}
The research of T. Nie and M. Rutkowski was supported by the DVC Research Bridging Support Grant {\it Pricing of American and game options in markets with frictions}. The work of T. Nie was supported by the National Natural Science Foundation of China (No. 11601285) and the Natural Science Foundation of Shandong Province (No. ZR2016AQ13).




\end{document}